\documentclass[a4paper,fleqn,usenatbib]{mnras}
\usepackage{mathptmx}

\usepackage[T1]{fontenc}

\usepackage{graphicx}   
\usepackage[fleqn]{amsmath}    
\usepackage{amssymb}    

\hypersetup{pdftitle=Star Disk Project: Paper II,
            pdfsubject=Astrophysics \& Cosmology,
            pdfkeywords={},
            pdfauthor=Kennedy et al.}

\newcommand{\msun}{\mathrm{M}_{\odot}}
\newcommand{\rsun}{\mathrm{R}_{\odot}}
\newcommand{\yr}{\mathrm{yr}}
\defcitealias{JustEtAl2012}{Paper~I}
\newcommand{\PaperI}{\citetalias{JustEtAl2012}}
\newcommand{\upvarphi}{\phi}

\begin{document}

\title[Star--disc interaction in galactic nuclei]{Star--disc interaction in galactic nuclei: orbits and rates of accreted stars}

\author[Kennedy et al.]
  {Gareth F. Kennedy$^1$\thanks{Corresponding author email: \href{mailto:gareth.f.kennedy@gmail.com}{gareth.f.kennedy@gmail.com}}, Yohai Meiron$^{2,1}$, Bekdaulet Shukirgaliyev$^{3,4}$\thanks{Fellow of the International Max Planck Research School for Astronomy and Cosmic Physics at the University of Heidelberg (IMPRS-HD)},\newauthor Taras Panamarev$^{3,4}$\footnotemark[2],
   Peter Berczik$^{1,4,5}$, Andreas Just$^4$, Rainer Spurzem$^{1,2,4}$\\
      $^1$ National Astronomical Observatories of China and Key Laboratory for Computational Astrophysics, Chinese Academy of Sciences,\\ 20A Datun Rd., Chaoyang District, 100012, Beijing, China\\
      $^2$ Kavli Institute for Astronomy and Astrophysics at Peking University, 5 Yiheyuan Rd., Haidian District, 100871, Beijing, China\\
      $^3$ Fesenkov Astrophysical Institute, Observatory 23, 050020 Almaty, Kazakhstan \\
      $^4$ Astronomisches Rechen-Institut, Zentrum f\"ur Astronomie, University of Heidelberg, M\"onchhofstrasse 12-14, 69120, Heidelberg, Germany\\
      $^5$ Main Astronomical Observatory, National Academy of Sciences of Ukraine, 27 Akademika Zabolotnoho St., 03680, Kyiv, Ukraine\\
}

\maketitle

\begin{abstract}
We examine the effect of an accretion disc on the orbits of stars in the central star cluster surrounding a central massive black hole by performing a suite of 39 high-accuracy direct $N$-body simulations using state-of-the art software and accelerator hardware, with particle numbers up to 128k. The primary focus is on the accretion rate of stars by the black hole (equivalent to their tidal disruption rate for black holes in the small to medium mass range) and the eccentricity distribution of these stars. Our simulations vary not only the particle number, but disc model (two models examined), spatial resolution at the centre (characterised by the numerical accretion radius) and softening length. The large parameter range and physically realistic modelling allow us for the first time to confidently extrapolate these results to real galactic centres. While in a real galactic centre both particle number and accretion radius differ by a few orders of magnitude from our models, which are constrained by numerical capability, we find that the stellar accretion rate converges for models with $N \geq 32$k. The eccentricity distribution of accreted stars, however, does not converge. We find that there are two competing effects at work when improving the resolution: larger particle number leads to a smaller fraction of stars accreted on nearly-circular orbits, while higher spatial resolution increases this fraction. We scale our simulations to some nearby galaxies and find that the expected boost in stellar accretion (or tidal disruption, which could be observed as X-ray flares) in the presence of a gas disc is about a factor of 10. Even with this boost, the accretion of mass from stars is still a factor of $\sim 100$ slower than the accretion of gas from the disc. Thus, it seems accretion of stars is not a major contributor to black hole mass growth.
\end{abstract}

\begin{keywords}
accretion, accretion discs -- celestial mechanics -- stellar dynamics -- galaxies: active -- galaxies: nuclei -- methods: numerical
\end{keywords}

\section{Introduction}\label{sec:INTRO}

Over the past decades, evidence has accumulated for the existence of massive black holes (MBHs) in the centre of most, possibly all, galaxies \citep[e.g.][]{KormendyRichstone1995, RichstoneEtAl1998, MagorrianEtAl1998}. These objects are thought to be the engine behind the extremely luminous quasars observed in the early universe. To explain these observations, the MBHs must have accreted material at a rate of $\sim 1\,\msun\,\yr^{-1}$ in order to reach their inferred masses of up to $\sim 10^{10}\,\msun$. It is generally believed that gas infall from large scales provides the necessary feeding for those MBHs to outshine their host galaxies and reach present day masses \citep[e.g.][]{Soltan1982, HaehneltRees1993, YuTremaine2002}, other models \citep{MK2005} argue that the necessary fuel is provided by stars captured by a gas disc as they plunge through it, and are then destroyed and accreted by the MBH.

Both scenarios require the presence of a gas disc. The idea that emission from a thin accretion disc (AD) is the source of quasar luminosity is consistent with the observed quasar spectral energy distribution in the optical and UV \citep{Shields1978, SunMalkan1989, Laor1990}. The disc forms as gas infalls into the central region of the galaxy and initially loses energy and settles into a rotating flattened structure. Angular momentum is gradually transported outwards by various processes, allowing matter to progressively move inwards, towards the centre of gravity. While falling deeper into the gravitational well, viscous processes heat up the now ionized gas and cause yet more energy and angular momentum to be thermally radiated out; this process also cools the disc leading to a steady state \citep{AbramowiczStraub2014}. When a gas parcel eventually reaches a region in the disc where a circular orbit is no longer possible due to General Relativistic effects, it will then plunge into the MBH. A model for such a stationary AD was developed by \citet{SS1973}, and this model has been the basis of many other models. The stationary AD model was extended by \citet{NT1973} to include General Relativistic effects, which are critically important close to the inner boundary of the AD. It is important to note that such an AD may extend to the parsec scale or larger, where the environment is dominated by stars, orders of magnitude more than the MBH's gravitational (Schwarzschild) radius.

When a gas disc is not present, a MBH does not normally emit any detectable electromagnetic radiation. However, when a star occasionally falls in from the surrounding cluster close enough to be disrupted, the material then heats up by dissipative processes and radiates some of its binding energy, further falling further into the MBH. This process is only possible when the tidal disruption radius (which depends on the stellar radius and the MBH mass) is larger than the Schwarzschild radius, which is the case for a Sun-like star when $M_\mathrm{bh} \lesssim 10^8\,\msun$. Because of the great Keplerian speed so deep in the MBH's potential well, the gaseous debris is hot enough to produce X-rays. Since these tidal disruption events (TDEs) \citep{Hills1975, Hills1976, FR1976} happen close to the MBH event horizon, they help constrain some of the MBH's parameters such as mass and spin. Tidal disruption of stars has famously been used as a boundary condition on the Boltzmann equation to determine the stellar mass distribution around MBHs \citep{BahcallWolf1976, BahcallWolf1977, DokuchaevOzernoi1977a, DokuchaevOzernoi1977b}, following an equivalent idea in plasma physics \citep{Gurevich1964}.

In the past decade, some TDE candidate events have been published \citep{Komossa2002, KomossaMerritt2008, WangEtAl2012}, with potentially many more hidden in data archives. There is only a handful of observations simply because such events are rare and often very distant \citep{Komossa2002}. \citet{MagorrianTremaine1999} calculated stellar TDE rates in simple galaxy models using loss-cone theory and predicted a rate of no more than about one event in $10^4$ years per galaxy (largely in agreement with \citealt{SyerUlmer1999}, who used similar methodology); this value is increased by 1--2 orders of magnitude when using a more recent MBH mass distribution and is roughly consistent with the inferred rate of soft X-ray outbursts from normally quiescent galactic nuclei \citep{DonleyEtAl2002}.

The loss-cone (also referred to as loss-wedge or loss-cylinder) is defined as the region in angular momentum and energy (or radius) phase space where stars have orbits that bring them inside their tidal disruption radius at pericentre. This region is quickly emptied on a time-scale of the orbital crossing time as stars directly plunge onto the MBH. The analysis of \citet{MagorrianTremaine1999} assumed the loss-cone is repopulated by 2-body relaxation and large-scale torques; but other processes can keep the loss-cone full for a longer time (1) accelerating angular momentum diffusion (reducing the effective relaxation time) and (2) enlarging the effective size of the loss-cone. For example, \citet{PeretsEtAl2007} used an extension of loss-cone theory and found that massive perturbers were far more efficient than 2-body relaxation in bringing stars within the zone of influence of the MBH. The presence of a second MBH in the galactic nucleus can potentially increase the rate dramatically \citep{ChenEtAl2011} or under some circumstances decrease it \citep{MerrittWang2005}. Likewise, rotation of the star cluster around the MBH \citep{FiestasSpurzem2010, FiestasEtAl2012} as well as the recoil of a post-coalescence MBH from a galactic nucleus \citep{GualandrisMerritt2009, LiEtAl2012} can have a dramatic influence on the TDE rate as compared to the simple loss-cone estimation. Finally, \citet[hereafter Paper I]{JustEtAl2012} explored the influence of both the gaseous disc and star cluster centred on the MBH on the TDE rates.

Research into the interaction of stars with an AD goes back to \citet{VilkoviskijBekbasarov1981} and \citet{Vilkoviskij1983}, who realised that the gas density at the inner regions was potentially high enough to induce substantial drag on stars crossing it, changing their orbits. More detailed studies that followed in the 1990s aimed to find an equilibrium state between the transport of stars into the central regions and their removal from the AD \citep{SyerEtAl1991, ArtymowiczEtAl1993, VokrouhlickyKaras1998, SubrKaras1999}. \citet{VilkoviskijCzerny2002} introduced a semi-analytic model based on the balance between relaxation and dissipation processes, noting in particular that 2-body relaxation both supplies the AD with stars and also has the opposite effect of elevating back previously trapped stars. \citet{SubrEtAl2004} used similar methodology with a more realistic disc model and found that the presence of an AD causes significant anisotropies such as depletion of counter-rotating stars and flattening of the stellar system. More progress was made by \citet{Rauch1995, Rauch1999} who for the first time applied numerical simulations to the problem. Notably, General Relativistic effects were included in his work. He showed that the orbits of stars quickly align with the AD's plane, as well as that a constant density core forms due to collisional effects.

These studies all made simplifying assumptions about the AD, adopting infinitely-thin static disc models (no feedback from stars), and about the star cluster, which was often assumed to be spherically symmetric and to start in equilibrium (neglecting relevant time-scales). {\PaperI} have relaxed some of these assumptions by conducting, for the first time, self-consistent direct \textit{N}-body simulations to investigate the interaction of a central star cluster surrounding a MBH with an AD. They used a stationary Keplerian rotating disc model (which was vertically extended rather than infinitely-thin) and resolved the dissipation of kinetic energy of stars passing through it (ram pressure effects), as well as star accretion by the MBH and 2-body relaxation. They showed that the rate of star capture by the AD and subsequent accretion onto the MBH was enhanced by a significant factor compared to the case where no disc was included. They also showed that, as predicted by \citet{VilkoviskijCzerny2002}, there is a stationary flow of stars within the disc toward the central MBH, which is determined by an equilibrium between diffusion by 2-body encounters and energy loss by the dissipative force from the AD.

In this work, we extend the models of \PaperI. Our new models are significantly more detailed, with larger number of particles (up to an order of magnitude), higher resolution in the central region (artificial accretion radius decreased by up to two orders of magnitude) and a more physical disc model (with disc thickness varying with radius rather than constant). This new simulation suite allows us to explore a large range of the free parameters, examine the statistical properties of the orbits of plunging stars and the relevant time-scales, as well as for the first time make an accurate scaling analysis and extrapolate our results to real galactic centers.

This paper is organised as follows: in Section~\ref{sec:method} we present the method for modelling star--star and star--gas interactions as well as the initial conditions for the star cluster and AD; in Section~\ref{sec:plunge} we closely examine the types of stellar orbits that result in accretion or tidal disruption (based on the distribution of final orbital eccentricity of accreted stars and mass growth rate of the MBH); in Section~\ref{sec:RES} we investigate the dependence on these on the simulation particle number resolution and accretion radius spatial resolution; in Section~\ref{sec:hrdisc} the effect of the structure of the gas disc is investigated by changing the disc height, while keeping the surface density profile fixed; in Section~\ref{sec:realgc} we discuss the scaling analysis of our simulation results and applicability to real galactic centres; finally, in Section~\ref{sec:CON} we present a discussion of our key results and conclusions.

\section{Methods}
\label{sec:method}

In this paper we present a suite of 39 new simulations, spanning a factor of 16 in particle number $N$ and up to $N=1.28\times10^5$ particles; all models are listed in Table~\ref{ModelParas}. This number is still quite small compared to the number of stars in a real galaxy centre, our \textit{star particles} are thus ``superparticles'' representing a group of stars. The computation times varied greatly between models, with one of the longest runs (\texttt{128k03r}) taking 180 days to complete two relaxation times ($\sim 1800$ H\'enon time units\footnote{\citet{Heggie2014} suggested naming the standard \textit{N}-body unit system (previously known just as \textit{N}-body units) after M. H\'enon to commemorate his work; we use this throughout this study. In this unit set, the total stellar mass $M_\mathrm{cl}$ and the gravitational constant $G$ are set to unity, and the total energy of the isolated stellar system is set to $-0.25$. For convenience, we will drop the unit name when unambiguous.}) using the $\upvarphi$\textsc{grape} code (see Section~\ref{sec:Stellar_component}) on 8 nodes, each equipped with one NVIDIA Tesla M2050 GPU.

\begin{table}
\caption{Model parameters for all simulation runs}
\label{ModelParas}
\begin{centering}
\begin{tabular}{lrrll}
\hline
\hline
Model name & $N$ & $r_\mathrm{acc}^*$ & $\epsilon_\mathrm{bh}$ & AD type\\
\hline
\texttt{008k01ng}     & 8k   &  1 & $10^{-6}$ & no gas\\
\texttt{008k03ng}     & 8k   &  3 & $10^{-6}$ & no gas\\
\texttt{008k03ns}     & 8k   &  3 & $0$       & $h_z$\\
\texttt{008k03}       & 8k   &  3 & $10^{-6}$ & $h_z$\\
\texttt{008k03rns}    & 8k   &  3 & $0$       & $h(R)$\\
\texttt{008k03r}      & 8k   &  3 & $10^{-6}$ & $h(R)$\\
\texttt{008k10ng}     & 8k   & 10 & $10^{-6}$ & no gas\\
\texttt{008k10}       & 8k   & 10 & $10^{-6}$ & $h_z$\\
\texttt{008k50ng}     & 8k   & 50 & $10^{-6}$ & no gas\\
\texttt{008k50}       & 8k   & 50 & $10^{-6}$ & $h_z$\\
\texttt{016k01ng}     & 16k  &  1 & $10^{-6}$ & no gas\\
\texttt{016k03ng}     & 16k  &  3 & $10^{-6}$ & no gas\\
\texttt{016k03}       & 16k  &  3 & $10^{-6}$ & $h_z$\\
\texttt{016k10ng}     & 16k  & 10 & $10^{-6}$ & no gas\\
\texttt{016k10}       & 16k  & 10 & $10^{-6}$ & $h_z$\\
\texttt{016k50ng}     & 16k  & 50 & $10^{-6}$ & no gas\\
\texttt{016k50}       & 16k  & 50 & $10^{-6}$ & $h_z$\\
\texttt{032k01ng}     & 32k  &  1 & $10^{-6}$ & no gas\\
\texttt{032k03ng}     & 32k  &  3 & $10^{-6}$ & no gas\\
\texttt{032k03ns}     & 32k  &  3 & $0$       & $h_z$\\
\texttt{032k03}       & 32k  &  3 & $10^{-6}$ & $h_z$\\
\texttt{032k03rns}    & 32k  &  3 & $0$       & $h(R)$\\
\texttt{032k03r}      & 32k  &  3 & $10^{-6}$ & $h(R)$\\
\texttt{032k10ng}     & 32k  & 10 & $10^{-6}$ & no gas\\
\texttt{032k10}       & 32k  & 10 & $10^{-6}$ & $h_z$\\
\texttt{032k50ng}     & 32k  & 50 & $10^{-6}$ & no gas\\
\texttt{032k50}       & 32k  & 50 & $10^{-6}$ & $h_z$\\
\texttt{064k01ng}     & 64k  &  1 & $10^{-6}$ & no gas\\
\texttt{064k03ng}     & 64k  &  3 & $10^{-6}$ & no gas\\
\texttt{064k10ng}     & 64k  & 10 & $10^{-6}$ & no gas\\
\texttt{064k50ng}     & 64k  & 50 & $10^{-6}$ & no gas\\
\texttt{128k01ng}     & 128k &  1 & $10^{-6}$ & no gas\\
\texttt{128k03ng}     & 128k &  3 & $10^{-6}$ & no gas\\
\texttt{128k03}       & 128k &  3 & $10^{-6}$ & $h_z$\\
\texttt{128k03r}      & 128k &  3 & $10^{-6}$ & $h(R)$\\
\texttt{128k10ng}     & 128k & 10 & $10^{-6}$ & no gas\\
\texttt{128k10}       & 128k & 10 & $10^{-6}$ & $h_z$\\
\texttt{128k50ng}     & 128k & 50 & $10^{-6}$ & no gas\\
\texttt{128k50}       & 128k & 50 & $10^{-6}$ & $h_z$\\
\hline
 \end{tabular}
\end{centering}
\par\medskip
\textbf{Notes.} All models completed at least two relaxation times, with many completing nine. Column 2 gives the number of particles $N$ used in each model where k stands for $10^3$ particles; Column 3 gives the spatial resolution at the centre, where $r_\mathrm{acc}^*$ is a shorthand for convenience and represents the numerical accretion radius in units of $10^{-4} R_\mathrm{d}$ (where $R_\mathrm{d}$ is the scale of the disc); Column 4 gives the softening length in H\'enon units of the star--MBH force or zero when there was no softening (\texttt{ns}), the star--star softening was $10^{-4}$ in all models; Column 5 gives the particular type of AD used for the mode, where $h_z$ means constant thickness, $h(R)$ means varying thickness as described in the text, and `no gas' are pure stellar dynamical models with no AD.
\end{table}

The spatial resolution is characterised by the \textit{numerical} accretion radius $r_\mathrm{acc}$, which is the distance from the MBH at which particles are said to be accreted by the MBH. When this happens in the simulation, the star particle disappears and its mass is added to the MBH. The value of $r_\mathrm{acc}$ is chosen as a compromise between physical realism and computational cost. While we improved the spatial resolution significantly from {\PaperI} by up to two orders of magnitude (the lowest resolution in this study was the highest in theirs), the smallest value of $r_\mathrm{acc}$ used here, when scaled to real galaxies, is still 2--3 orders of magnitudes larger than the tidal radius for a Sun-like star (the motivation here is to ensure that the $\upvarphi$\textsc{grape} code can treat star accretion reliably). In Section~\ref{sec:realgc} we discuss the implications of this choice. For convenience we define the quantity $r_\mathrm{acc}^* \equiv r_\mathrm{acc}/(10^{-4}R_\mathrm{d})$, where $R_\mathrm{d}$ is the scale of the disc as explained below.

In all models listed in Table~\ref{ModelParas}, the star cluster mass $M_\mathrm{cl}$ is set to unity. The MBH is fixed at the centre or the coordinate system and its mass is $M_\mathrm{bh}=0.1$.

In the following subsections, we give a brief summary of the methods used in {\PaperI}, focusing on the changes made in the $\upvarphi$\textsc{grape} code since then in terms of the disc model, stellar component and the interaction between the stars and the disc.

\subsection{Accretion disc component}\label{sec:discmodel}
We have two models for the AD, denoted $h_z$ (constant thickness) and $h(R)$ (varying thickness); both correspond to an axisymmetric thin disc inspired by \citet{SS1973} and \citet{NT1973}. The gas density is given by
\begin{eqnarray}
\rho_\mathrm{g}(R,z) & = & \frac{2-p}{2\pi \sqrt{2 \pi}} \frac{M_\mathrm{d}}{h R_\mathrm{d}^2} \left( \frac{R}{R_\mathrm{d}} \right)^{-p}\nonumber\\
 & & \exp \left[ -\beta_s \left( \frac{R}{R_\mathrm{d}} \right)^s \right] \exp \left( \frac{-z^2}{2 h^2} \right),
\label{gasdensity}
\end{eqnarray}
where $p=3/4$ is the structure power-law corresponding to the outer region of standard thin disc model, $R$ is the radial distance from the MBH, $z$ is the vertical distance from the disc plane, $R_\mathrm{d} = 0.22$ is the radial extent of the disc; its value is determined such that the enclosed stellar mass within $R_\mathrm{d}$ equals the MBH mass at the beginning of the simulation (after letting the stellar system relax for $\sim 5$ crossing times in the MBH potential in order to get a dynamical equilibrium). The parameters $s=4$ and $\beta_s=0.7$ are associated with the smoothness of the outer cutoff of the disc (introduced for numerical reasons) and $h$ is the scale height (note that {\PaperI} defined $h$ differently, as a dimensionless ratio). The total disc mass is fixed to be $M_\mathrm{d} = 0.01$, with the numerical minimum and maximum radii given by $r_\mathrm{acc}$ and 1, respectively (but the density drops sharply around $R=R_\mathrm{d}$). The gas in the disc is taken to have a Keplerian rotation profile.

The two disc models differ by the vertical disc structure. The first model, denoted by $h_z$, has a constant scale height $h_z = 10^{-3} R_\mathrm{d}$. This corresponds to a self-gravitating isothermal profile, which is sufficiently accurate for large radii, and was thus adopted by {\PaperI} as they could not resolve the innermost part of the disc.

The second disc model modifies the former by using a simple linear relation for the scale height in the inner region. This is based on the fact that an AD in hydrostatic equilibrium with in the MBH gravitational field curves like $R^{9/8}$, which is reasonably similar to constant slope. The disc height is thus proportional to $R$ up to a radius where the disc is vertically self-gravitating at $R = R_\mathrm{sg}$, beyond which it is given by $h=h_z$, i.e.
\begin{equation}
h(R)=\begin{cases}
\frac{R}{R_\mathrm{sg}}h_z & R \leq R_\mathrm{sg}\\
h_z & R > R_\mathrm{sg}.
\end{cases}\label{eq:hR}
\end{equation}

The transition between the two regions can be estimated by equating the vertical component of the spherically symmetric force from the MBH at $z=h_z$ with the vertical self-gravitation of a thin disc above the AD:
\begin{equation}
\frac{GM_\mathrm{bh}h_z}{R_\mathrm{sg}^3} = 2\pi G \Sigma(R_\mathrm{sg}),\label{eq:find_Rsg}
\end{equation}
where $\Sigma(R)$ is the surface density of the disc which is the integral $\mathrm{d}z$ of equation~(\ref{gasdensity}); as only the last exponent is a function of $z$, the result is a simple power-law in $R$ with the same index $p$ and an exponential cutoff.

We note that both sides of equation~(\ref{eq:find_Rsg}) are approximate; we are only interested in an order of magnitude estimate of $R_\mathrm{sg}$ since the model for the AD structure is very rough anyway. On the left hand side, we only keep the first order term in a Taylor expansion of the vertical force component, which is linear in $z$ (the exact result is only $\sim 15$ per cent smaller). On the right hand side, the expression is the standard thin disc approximation solving the vertical Poisson equation while neglecting the radial derivatives \citep[for more information see][]{BinneyTremaine2008}. The solution is
\begin{equation}
R_\mathrm{sg}^{3-p} = \frac{1}{2-p} \frac{M_\mathrm{bh}}{M_\mathrm{d}} h_z R_\mathrm{d}^{2-p} .\label{rsg}
\end{equation}
For our choice of parameters, we get $R_\mathrm{sg} \approx 0.026$ which gives half-opening angle of $0.5^\circ$. The effects of changing the disc height profile on the results are examined in Section~\ref{sec:hrdisc}.

\subsection{Stellar component}
\label{sec:Stellar_component}

The stellar cluster initially had \citet{Plummer1911} density profile; it was allowed to dynamically evolve in the presence of the MBH for $\sim 5$ crossing times before $t=0$ and the AD is added. The numerical integration is done with a version of the $\upvarphi$\textsc{grape} code \citep{HarfstEtAl2007} which has been modified to include the effect of an AD and a central MBH. It is a direct \textit{N}-body code based on the fourth-order Hermite integration scheme. It is a parallel code that also uses accelerator technology such as GPUs to calculate mutual gravitational interactions between star particles. The $\upvarphi$\textsc{grape} code has been used in {\PaperI} as well as many other publications about galactic centres \citep[e.g.][and references thereafter]{BerczikEtAl2005}.

The $\upvarphi$\textsc{grape} code suits this investigation due to its computational speed and simplicity, which allows it to be easily modified. The version used here does not include stellar evolution, star formation, nor a mass function. There is no special treatment of two (or more) body encounters; i.e. no regularisation for close encounters. Neglecting these factors is justified here since the particle resolution is not high enough to model individual stars.

Indeed, a star particle does not actually represent a single star, but rather a group of stars. The number of stars in such a so-called \textit{superparticle} varies wildly between models and the galaxies they are scaled to; in the best case (models with largest $N$) ranges between a few hundreds (small galaxies) to hundreds of thousands (giant galaxies). This means that softening of the gravitational force is needed between the star particles, and we used $\epsilon = 10^{-4}$ in all models.

An additional softening of the same form is used in the interaction between a star particle and the MBH; as seen in Table~\ref{ModelParas}, most of the models use $\epsilon_\mathrm{bh} = 10^{-6}$. This value is not expected to significantly affect the results as the softening is smaller than all $r_\mathrm{acc}$ values. Using softening for the MBH interaction does complicate the orbital elements for the stars at the point of accretion. By correcting the MBH-star distance by $|\boldsymbol{r}|^2 \rightarrow |\boldsymbol{r}|^2+\epsilon_\mathrm{bh}^2$ in the angular momentum and binding energy, the orbital elements can be accurately computed. For models with $\epsilon_\mathrm{bh}>0$ the corrected values for eccentricity and semi-major axis are used in all presented results. The effect of softening on the final orbits of accreted stars and the growth of the MBH was tested by setting $\epsilon_\mathrm{bh}=0$ in four of the models in Table~\ref{ModelParas}. We examined the effects of softening length and saw very little difference between the softened and unsoftened models. The eccentricity distribution is slightly different, with stars being accreted on more eccentric orbits for the unsoftened models.

\subsection{Star--disc interaction}
\label{sec:SDI}

As a star particle passes through the accretion disc, a dissipative force is applied against its motion. Once its orbit lies within the disc (meaning that the apocentre $< R_\mathrm{d}$ and the highest point of the orbit above the plane of the disk $< h_z$), the particle is considered to be captured into the disc. If the distance of a star particle to the MBH comes within $r_\mathrm{acc}$, it is considered accreted by the MBH and is removed from the simulation.

The dissipative force that an individual star experiences while crossing the disc comes from ram pressure and originates at the bow shock in front of the star. It is proportional to the geometric cross section and local gas density, and is quadratic in velocity. It is given by the drag equation:
\begin{equation}
\boldsymbol{F}_\mathrm{drag} = - Q_\mathrm{d} \pi r_\star^2 \rho_\mathrm{g}(R,z) \left| \boldsymbol{V}_\mathrm{rel} \right| \boldsymbol{V}_\mathrm{rel},\label{eq:fdrag_star}
\end{equation}
where $Q_\mathrm{d}$ is the drag coefficient, $\rho_\mathrm{g}$ is the local gas density (equation \ref{gasdensity}), $r_\star^2$ is the stellar radius and $\boldsymbol{V}_\mathrm{rel}$ is the relative velocity between the star and the gas. The drag coefficient $Q_\mathrm{d}$ incorporates much of the uncertainty and can be estimated from the shock conditions \citep{CourantFriedrichs1948}; we use $Q_\mathrm{d}=5$. This prescription is only appropriate when the star is supersonic with respect to the disc and ignores gravitational focusing; see section 2.2 in {\PaperI} for more details.

Since a star particle does not represent a single Sun-like star but rather a group of stars, equation~(\ref{eq:fdrag_star}) has to be properly scaled in order to be used in a simulation. We do this by introducing an effective dissipative parameter,
\begin{equation}
Q_\mathrm{tot}(N) \equiv Q_\mathrm{d} N \left( \frac{r_\star}{R_\mathrm{d}}\right)^2.\label{Qtotdef}
\end{equation}
Substituting into equation~(\ref{eq:fdrag_star}), we get
\begin{equation}
\boldsymbol{a}_\mathrm{drag} = - Q_\mathrm{tot} \frac{\pi R_\mathrm{d}^2 \rho_\mathrm{g}(R,z)}{M_\mathrm{cl}} \left| \boldsymbol{V}_\mathrm{rel} \right| \boldsymbol{V}_\mathrm{rel}.\label{fdrag}
\end{equation}
Note that equation~(\ref{fdrag}) contains only \textit{global} quantities related to the entire cluster, and thus easily expressed in H\'enon units. Despite that, $Q_\mathrm{tot}$ is still very much a `real world' quantity rather than a simulation one (i.e. the $N$ in equation~\ref{Qtotdef} is $N_\mathrm{real}$, the number of stars in a galactic centre, not the number $N_\mathrm{sim}$ of particles in the model). Before we can use equation~(\ref{fdrag}) in the simulation, $Q_\mathrm{tot}$ has to be treated carefully because it introduces a new physical time-scale.

The main concern when modelling a system with superparticles rather than particles representing individual stars is that the relaxation time (which depends on the number of particles) is much shorter, while the dynamical time (only depends on the mass) is the same. The following expression is the relaxation time at the half mass radius \citep{Spitzer1987}:
\begin{eqnarray}
t_\mathrm{rx}(N) & = & \frac{0.14 N}{\ln \left( 0.4 N \right)} t_\mathrm{dyn}\label{trelax}\\
t_\mathrm{dyn}   & = & \left( \frac{r_\mathrm{h}^3}{G M_\mathrm{cl}} \right)^{1/2}.\label{tdyn}
\end{eqnarray}
As the half-mass radius at the beginning of the simulation (after letting the system evolve with no AD for 5 crossing times) is measured to be $r_\mathrm{h} = 0.66$, the dynamical time of the system is $\sim 0.54$ H\'enon time units, and the relaxation time ranges between 74 ($N=8$k) to 886 ($N=128$k) time units. 

The way to deal with the fact that $t_\mathrm{rx}$ depends on the number of particles is to present the results in terms of relaxation times, and then scale to physical time units when comparing with a real galactic centre. However, this cannot work if there are more physical processes associated with their own time-scales. The accretion rate of stars to the MBH is a secular process and is driven by two physical mechanisms: relaxation and dissipation. Two-body relaxation scatters stars into the empty part of the loss cone, which are then accreted quickly on high eccentricity orbits, its characteristic timescale increases with particle number $N$ (equation \ref{trelax}). The second process is energy loss by the drag force with the gas disk, which depends on the strength of the drag force at each crossing and the inverse of the orbital time (orbital frequency). It scales with the dissipation timescale as described in {\PaperI} and is independent of the particle number.

In order to correctly reproduce the accretion process, we need to remove the different $N$-dependences of the two mechanisms, i.e. we need to hold the ratio of dissipation and relaxation time constant. We achieve this by rescaling the
parameter $Q_\textrm{tot}$ as function of $N_\textrm{sim}$ particles by
\begin{equation}
Q_\mathrm{tot}(N_\mathrm{sim}) = \frac{t_\mathrm{rx}(N_\mathrm{real})}{t_\mathrm{rx}(N_\mathrm{sim})} Q_\mathrm{tot}(N_\mathrm{real}),
\label{QtotNrel}
\end{equation}
where $N_\mathrm{real}$ is the number of stars in a galactic centre with  and an effective dissipative parameter $Q_\mathrm{tot}(N_\mathrm{real})$. In order to calibrate all our models using equation~(\ref{QtotNrel}), some real world values, i.e. $N_\mathrm{real}$ and $Q_\mathrm{tot}(N_\mathrm{real})$, have to be chosen. If we consider an AD of $R_\mathrm{d}=10\,\mathrm{pc}$ surrounded by approximately $2 \times 10^9$ Sun-like stars, equation~(\ref{Qtotdef}) gives $Q_\mathrm{tot} \approx 5\times10^{-8}$. Using a combination of  equations~(\ref{QtotNrel}) and (\ref{trelax}) to scale this to arbitrary $N$, we set
\begin{equation}
Q_\mathrm{tot}(N) \approx 5.42 \ln(0.4N)/N.
\end{equation}

This method of treating the star--disc interaction does not include: feedback onto the disc from stellar crossings; gravity force from the disc mass, acting on either the stars or the gas; and finally the contribution of stellar winds to the gas disc via mass or energy. These effects will be explored in future work.

\section{Star plunging}
\label{sec:plunge}

\subsection{Plunge types}
\label{starplunge}

The drag force due to the gas disc (equation~\ref{fdrag}) will act to dissipate the orbital binding energy of a star particle. This will shrink the orbit and slowly align it to the plane of the disc. When the orbit lies completely within the disc it can be said to be captured into the disc. However, this process can be interrupted at any time if the star comes within $r_\mathrm{acc}$ of the MBH.

This then leads to three paths for a star to be accreted onto the MBH: (1) \textit{disc capture} of particles with low eccentricity and low relative inclination between the orbital plane and the accretion disc, (2) \textit{gas assisted accretion} of particles with moderate eccentricity, characterised by a uniform inclination distribution, and (3) \textit{direct accretion} onto the MBH of particles on nearly radial orbits (also characterised by a uniform inclination distribution). The three paths correspond to $e_\mathrm{acc}\sim0$, $<1$, and $\sim 1$ respectively, where $e_\mathrm{acc}$ is the particle's orbital eccentricity at the time of accretion. These three broad \textit{plunge types} are thus categorised by the eccentricity at the time of accretion.

There are five basic phases that a star can go through (in each of the above categories): (1) scattering, (2) slow decay, (3) fast decay, (4) disc migration, and (5) radial infall. Note that not every accreted star will go through all of these phases. For example, before a gas assisted accretion event, a star will go through the scattering and possibly some of the slow migration phase before it is almost radially accreted onto the MBH, while the radial fall phase is unique to the radially accreted type. The scattering phase is when the star is far from the MBH, almost unaware of the disc, and any change in binding energy and angular momentum is caused by interactions between stars. Once the star is scattered close enough into the disc that disc crossings can significantly alter the binding energy, then it is in the slow decay phase. If the star is aligned with the disc when the radius is within $\sim 10^{-3}$ length units, then its orbit will rapidly decay in $\lesssim 10$ time units. If the star becomes completely aligned with the disc on a nearly circular orbit, it enters the final phase where it migrates with the gas disc until eventually falling inside $r_\mathrm{acc}$, also in $\lesssim 10$ time units. We use the word \textit{plunge} to refer to all phases an accreted star goes through after the scattering phase, and thus the plunge type as defined above refers to path to accretion a star has taken.

\begin{figure*}
\includegraphics[width=1\textwidth]{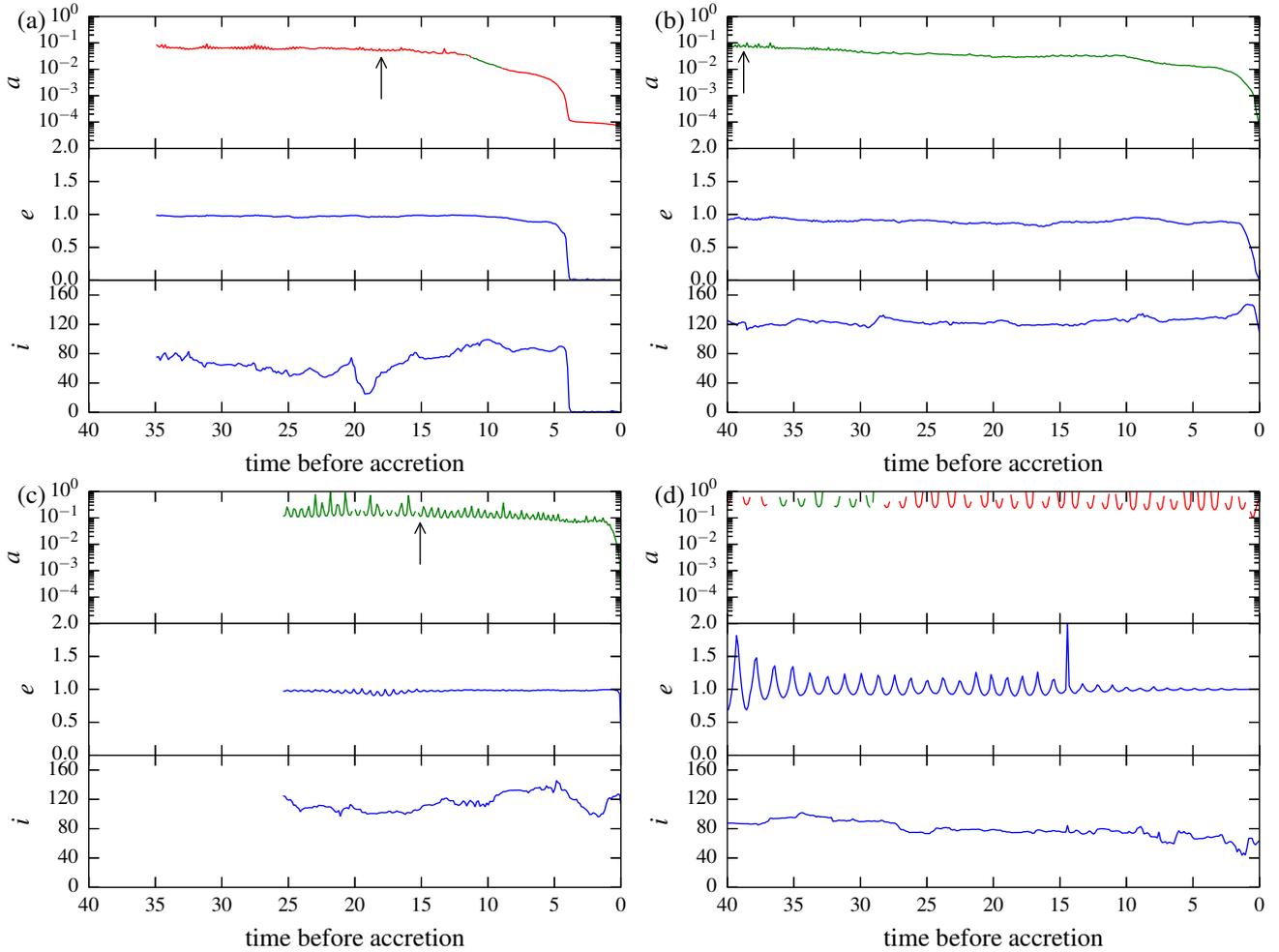}
\caption{Semi-major axis, eccentricity and inclination of a few sample orbits for the three types described in Section \ref{starplunge}. The semi-major axis is plotted in red when the orbits is prograde ($i<90^\circ$) and green when retrograde. Panel (a) shows a disc captured star with $e_\mathrm{acc} \sim 0$ and low inclination prograde orbit, (b) shows a star captured into the disc on a retrograde orbit, (c) shows a gas assisted accretion $e_\mathrm{acc} < 1$ where the inclination distribution for those orbits is uniform, (d) shows a direct accretion onto the MBH with $e_\mathrm{acc} \sim 1$ and also from a uniform inclination distribution. Black arrows indicate where the decay phase begins (see text). Panel (d) shows a direct capture, and thus has no decay phase; the gaps in the semi-major axis occur when the orbit is instantaneously unbound.}
\label{FigPlunges}
\end{figure*}

Fig.~\ref{FigPlunges} shows the semi-major axis, eccentricity and inclination as a function of time \textit{before} the star is accreted (i.e. time increases to the right of the horizontal axis) for selected stars in the above three categories. The orbital elements $a$ and $e$ are calculated from the particle's orbital energy and angular momentum under the assumption of a Keplerian orbit, they are thus effective quantities rather than geometric ones. The orbits are not perfectly Keplerian due to the self-gravity of the stellar system, however the relevant orbits are deep enough in the MBH's potential well to justify this approximation (as noted earlier, $r_\mathrm{inf}=R_\mathrm{d}=0.22$). The ``wiggling'' seen especially in panels (c) and (d) is due to the fact that these orbits are on the outskirts of the MBH's sphere of influence, where this approximation is poor. The black arrows show an estimate of the beginning of the decay phase. This time is characterised by a consistent decrease in the binding energy of the orbit. The method for determining this point is involved and the details are discussed in the next Section.

The first category of disc capture can be further divided into two subcategories based on the orbit's orientation relative to the disc's angular momentum. An example of a star captured into the disc and later accreted onto the MBH from a prograde orbit is shown in Fig.~\ref{FigPlunges} (a), while an example of an accretion from a retrograde orbit is shown in Fig.~\ref{FigPlunges} (b). For both prograde and retrograde disc capture stars, the star becomes aligned with the disc and the orbit rapidly circularises once it lies within the disc. Note that the semi-major axis of the particle on the retrograde orbit decays much more rapidly than the prograde one due to head wind effect.

For a particle orbiting within the disc (either prograde or retrograde), by the time it reaches $r_\mathrm{acc}$, the orbital eccentricity has decayed to $\sim 0$. The details of the motion of the particle once it is captured into the disc are complicated and not accurately modelled here. This disc migration phase is similar to a Type I migration in planetary dynamics, i.e. large mass ratio between the captured star and the gas disc, so no gap opening is expected \citep[cf.][]{BaruteauEtAl2011, KocsisEtAl2011, McKernanEtAl2011, BellovaryEtAl2015}. The final migrating phase is slow in our simulations since the relative velocity between the star and the gas goes to zero, and so does the force in equation~(\ref{fdrag}). In nature, this decay will be much faster since the gas will have an inward radial velocity component. This level of realism is not included in our models since we cannot resolve a single star, nor do we model all of the microphysics.

The second type (gas assisted accretion) is shown in Fig.~\ref{FigPlunges} (c). This example starts with a large relative inclination with respect to the disc plane (and retrograde orientation) such that the star particle is accreted onto the MBH before there is sufficient time for the eccentricity to decay. This lack of time to circularise is the only difference between this type and the first type where stars are accreted after spending time within the gas disc itself. In this case the particle quickly begins to plunge through the disc (black arrow) and enters the slow decay phase where it steadily loses binding energy until it comes within the accretion radius. We found many similar examples in the simulation results that showed a long scattering phase (see below) before the slow decay. These can potentially last multiple relaxation times until the star crosses the disc close enough for it to enter into the slow decay phase. The statistics of how deep (i.e. close to the MBH) a star must cross the disc to begin the decay phase is discussed in Section~\ref{plungestats}.

The third type is a direct accretion of the star particle on a nearly radial orbit onto the MBH. These events are caused by an interaction between two distant stars, one of which is sling shot into the loss-cone of the MBH. The example particle shown in Fig.~\ref{FigPlunges} (d) comes from a radial orbit ($e_\mathrm{acc} \sim 1$) at a large distance from the MBH. This type has little to no interaction with the gas disc, and so has no arrow as there is no decay phase. The accretion happens on an orbital time which is not visible in the plot. Note that the gaps in the semi-major axis versus time are due to the particle being on hyperbolic (unbound) orbits and thus negative semi-major axis. This process is well studied and for an estimation of the mass growth of the MBH based on loss-cone filling see Section~\ref{sec:losscone}. The scattering time-scale before the orbital decay is largely variable and can be of the order of a relaxation time, whereas the time-scale for the final phase for direct accretion is just the free-fall time to the MBH.

\subsection{Plunge statistics}
\label{plungestats}

The aim of this Section is twofold: (1) to estimate the plunge time-scales and (2) to estimate the distance from the MBH where the slow decay phase begins. This distance is used to estimate an effective radius for the disc. Since not all stars coming within this radius will instantly be accreted, this radius cannot be converted into an effective cross section, like the tidal radius can for loss-cone calculations in stellar systems (see Section~\ref{sec:losscone}).

Before proceeding, let us note that the beginning of the plunge is the end of the scattering phase, and could mean either the beginning of the decay (types 1 and 2) or beginning of the radial infall (type 3). The latter is not so relevant in the context of the star-disc interaction as type 3 plunge is a 2-body effect that has nothing to do with the disc.

To define the time of beginning of the plunge for types 1 and 2, one first needs to find the peak of the fast orbital decay phase, i.e. the most recent minimum in $\mathrm{d}a/\mathrm{d}t$ while the star particle is bound to the MBH ($a>0$) and before the star is accreted. Then, the time derivative $\mathrm{d}a/\mathrm{d}t$ is followed back in time to find the next local maximum in semi-major axis, this indicates the previous increase in semi-major axis. Since the semi-major axis can only be increased by an interaction with another star, this marks the end of the scattering phase, and is defined as the beginning of the plunge. The time derivative is calculated from $a(t)$ after being smoothed using a moving average with a window of 1 time unit to remove spurious low amplitude oscillations during the plunge caused by scattering events and deviations from the assumed Keplerian potential. Examples of the beginning of plunges for those two types are shown as the black arrows in Fig.~\ref{FigPlunges} (a--c).

This procedure was applied to all accretion events from the most realistic gas disc model from Table~\ref{ModelParas}, namely \texttt{128k03r}. After two relaxation times, the total number of accreted stars was 14657, of which 19.3 per cent had $e_\mathrm{acc} > 0.8$ (most of which direct accretion onto the MBH and so had no orbital decay phase), 18.6 per cent had eccentricities in the range $0.1 < e_\mathrm{acc} < 0.8$ and 62.1 per cent had $e_\mathrm{acc} < 0.1$, typically very close to zero. For more details on the orbital distribution of accreted stars from this model see Section~\ref{sec:hrdisc}.

\begin{figure}
\begin{centering}
\includegraphics[width=\columnwidth]{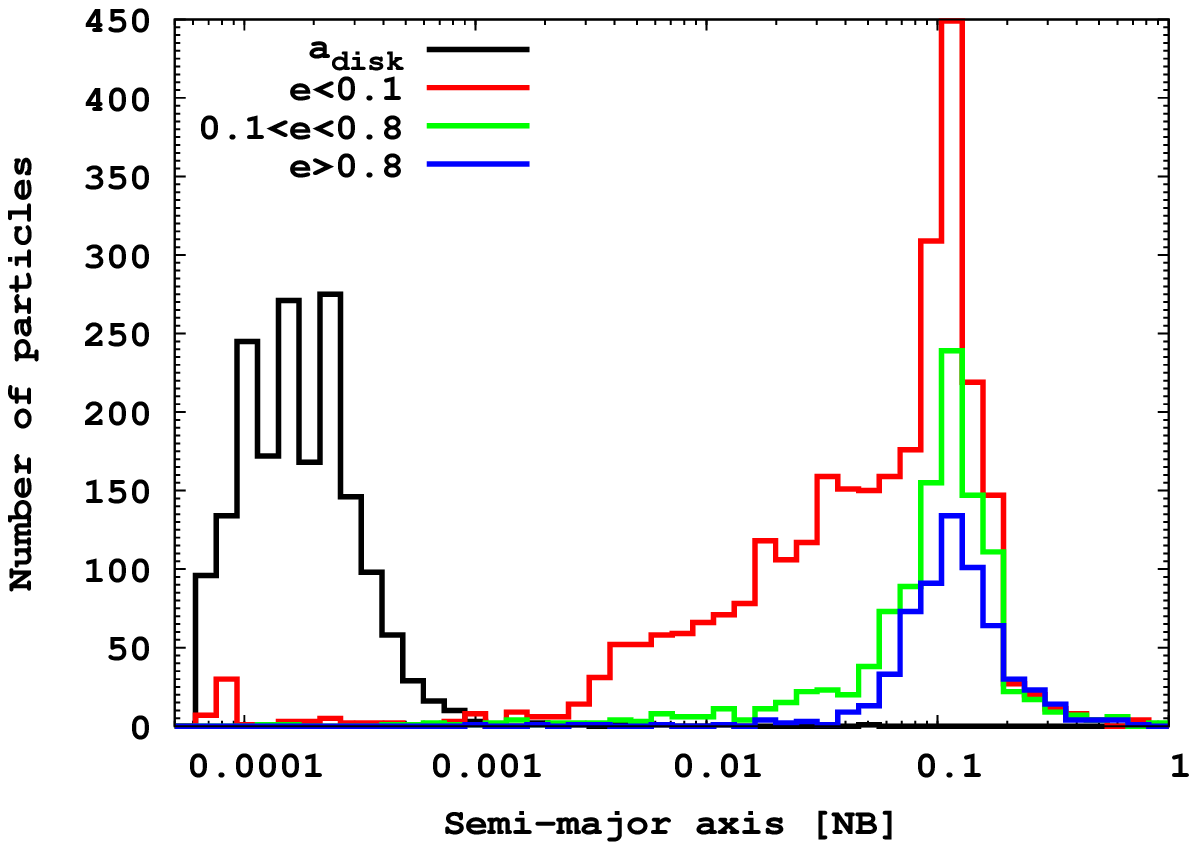}
\par\end{centering}
\caption{Distribution of semi-major axis at the \textit{beginning} of the plunge in model \texttt{128k03r} where 14657 particles plunged during about two relaxation times, divided into three groups according to \textit{final} eccentricity. Low $e_\mathrm{acc}$ tend to begin their plunge from lower semi-major axis than higher eccentricity particles. Particles are removed from the simulation at $r_\mathrm{acc} = 6.6 \times 10^{-5}$ H\'enon length units. $a_\mathrm{disc}$ is the distribution of radii where the particle had an orbit completely inside the disc. This shows that stars go into the disc before they are accreted at the accretion radius.}
\label{Plunge:semi}
\end{figure}

The distribution of semi-major axes at the \textit{beginning} of each plunge is shown in Fig.~\ref{Plunge:semi} for all accreted stars broken down to three groups according to their \textit{final} eccentricities $e_\mathrm{acc}$. For plunge types 1 and 2, the plotted quantity is the semi-major axis at the time of the beginning of the plunge as defined above; for type 3, the value shown is the semi-major axis at the time of accretion for bound (elliptical) orbits only, which has the same value as in the beginning of the plunge because in this case the plunge (radial infall) is very quick. One difference between the three distributions is that stars with $e_\mathrm{acc}<0.1$ often begin their plunge in a much denser regions of the disc, as seen by the semi-major axis spread closer to the MBH.

\begin{figure}
\begin{centering}
\includegraphics[width=\columnwidth]{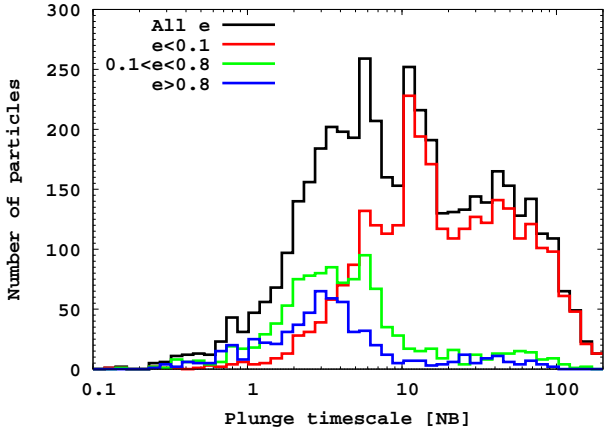}
\par\end{centering}
\caption{Distribution of plunge time-scale (same model and eccentricity grouping as in Fig.~\ref{Plunge:semi}). Among the particles accreted through the disc (low and moderate $e_\mathrm{acc}$), it is interesting to note that that the particles with low eccentricity take longer to be accreted. This is due to the final stage taking longer than it would in reality, since the particles end up moving with the gas disc velocity, which in our model has no radial component. For comparison, the half-mass dynamical time is $\sim 0.54$ H\'enon time units while the relaxation time for $N=128\mathrm{k}$ is $\sim 886$ time units.}
\label{Plunge:time}
\end{figure}

Another difference is in the plunge time-scales, as shown in Fig.~\ref{Plunge:time} where it is clear that while stars with low eccentricities at accretion begin the plunge from smaller semi-major axes, it takes longer for the semi-major axis to decay. Two examples of these plunges with $e_\mathrm{acc} \sim 0$ were given in Fig.~\ref{FigPlunges}, in both cases the particle has time to circularise before the semi-major axis decays and brings the star into the MBH. This process also aligns them to the disc and orbit in either a prograde or retrograde sense, this is examined further in Section~\ref{sec:hrdisc}. It is worth noting that the plunge time-scale can take up to $\sim 0.1$ times the relaxation time for stars that will become trapped in the disc (low $e_\mathrm{acc}$).

\begin{figure}
\begin{centering}
\includegraphics[width=\columnwidth]{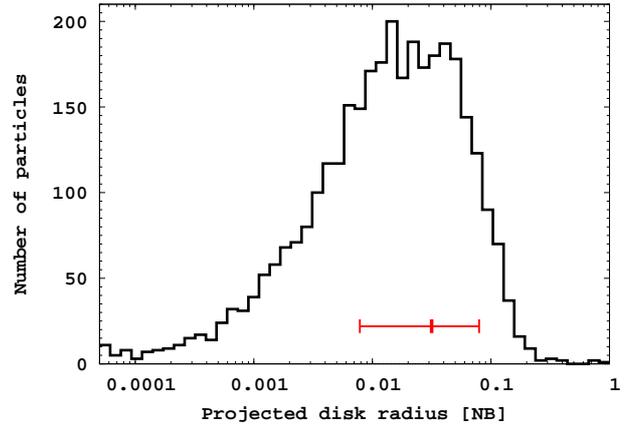}
\par\end{centering}
\caption{Distribution of the projected disc crossing radius, where the orbit crosses the disc just as the particle begins its plunge through the disc and onto the MBH (same model as Fig.~\ref{Plunge:semi}). The red horizontal bar indicates the median and standard deviation of the distribution, giving a characteristic location in the disc where most stars begin the plunge of $R_\mathrm{eff} = 0.032$.}
\label{Plunge:rdisc}
\end{figure}

To get a sense of the effective radius of the disc, the semi-major axis (the distribution of which is shown in Fig.~\ref{Plunge:semi}) is not the most important quantity to look at. Instead, we examine the \textit{projected disc crossing radius}, which is the radius where the orbit of the particle at the time it began its plunge crosses the disc; in this case only plunge types 1 and 2 are considered, since type 3 has no interaction with the disc. This value will be between the pericentre and the semi-minor axis. An approximation to the disc crossing distance is given by the distance from the MBH when the true anomaly is $\pm \pi/2$, i.e. $a(1-e^2)$, the distribution of which is shown in Fig.~\ref{Plunge:rdisc}.

The median value of the distribution given in Fig.~\ref{Plunge:rdisc} is used as an effective radius of the disc $R_\mathrm{eff} = 0.032$, which corresponds to an enclosed disc mass of $\sim 42$ per cent. This value is an estimate due to the ambiguity of defining the beginning of the plunge itself. Furthermore, the enclosed mass definition is not likely to be generally true for any kind of an AD, and investigation of a larger range of disc models and mass ratios is needed. This radius cannot be used to calculate a cross section as not every particle that crosses the disc within this distance will be eventually accreted.

A better way of thinking about the effective radius is as a divider between potential plungers ($R<R_\mathrm{eff}$) and stars that will be sped up (in the prograde direction) without being accreted ($R>R_\mathrm{eff}$). The latter category of stars will cross the disc and have their velocities changed by the gas velocity, which will induce a rotation in the star cluster over time. This effect of the accretion disc on the star cluster is examined in more detail in a future work.

\begin{figure}
\includegraphics[width=1\columnwidth]{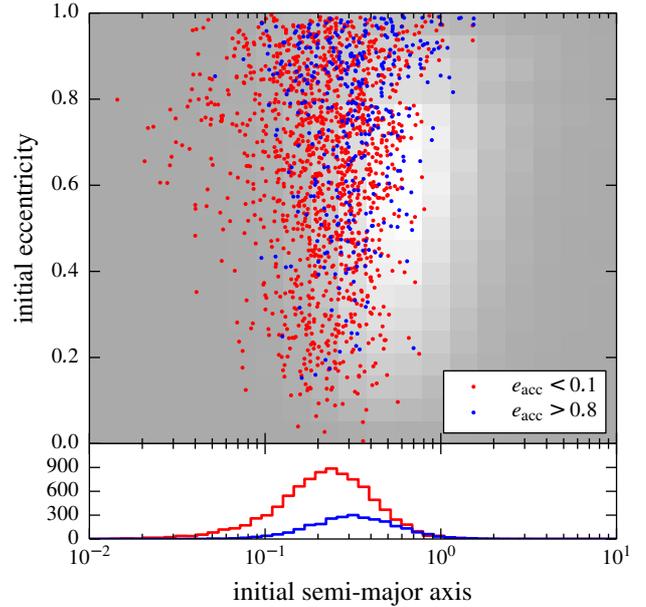}
\caption{Distribution of the \textit{initial} semi-major axis and eccentricities of the accreted particles grouped according to the \textit{final eccentricity} (plunge type). Cf. Fig. \ref{Plunge:semi} that show the semi-major axis at the beginning of the plunge. The blue points represent the initial conditions of particles that went through type 3 plunge ($e_\mathrm{acc}>0.8$) while the red points are particles that go through type 1 plunge ($e_\mathrm{acc}<0.1$); only one in seven points is shown in order not to overcrowd the figure, and type 2 has an intermediate distribution which we do not show for the same reason. The background color represents the initial conditions of all particles.
\label{fig:InitialScatter}}
\end{figure}

In contrast with the above figures that present quantities measured at the beginning of the plunge, Fig. \ref{fig:InitialScatter} shows the distribution of the \textit{initial} semi-major axis and eccentricities of the accreted particles grouped according to the \textit{final eccentricity} (plunge type). 

The blue points represent the initial conditions of particles that went through type 3 plunge ($e_\mathrm{acc}>0.8$) while the red points are particles that go through type 1 plunge ($e_\mathrm{acc}<0.1$); only one in seven points is shown in order not to overcrowd the figure, and type 2 has an intermediate distribution which we do not show for the same reason. The background color represents the initial conditions of all particles (rather than just accreted ones). When we discussed the orbital elements previously, it was in the context of the plunge, which occurs within the sphere of influence of the MBH (otherwise the gas is just too sparse to initiate orbital decay), so the Keplerian approximation to calculate the orbital element was valid. This is not the case here, and we calculated the geometric semi-major axis and eccentricity from the peri- and apocentres by integrating each orbit in the smoothed potential using the ETICS-SCF code \citep{MeironEtAl2014}
\begin{eqnarray}
a_\mathrm{geo} & = & (r_\mathrm{apo}+r_\mathrm{peri})/2\\
e_\mathrm{geo} & = & (r_\mathrm{apo}-r_\mathrm{peri})/2a.
\end{eqnarray}

The two histograms in the bottom panel show the initial semi-major axis distribution of these types, and are roughly log-normal distributions of similar width (cf. the distribution of semi-major axes at the beginning of the plunge in Fig. \ref{Plunge:semi}); the red histogram contains more than three times the number of particles in the blue one, which represents the ratio between type 1 and type 3 plunges (as noted above) this could be understood by looking at the disc as a ``bigger target'' (however as noted before, $R_\mathrm{eff}$ is not equivalent to a geometric cross section).

Particles which started at a semi-major axis of $\gtrsim 1$ are reasonably ``safe'' from being accreted onto the MBH: a little less than 0.5 per cent of particles with $a_\mathrm{init}>1$ (and any eccentricity) were accreted after two relaxation times, compared to 15 per cent for the complementary set. This is due to the fact that 2-body relaxation at the outskirts is really much slower than at the half-mass radius, where the relaxation time is measured. Thus, particles with very large initial $a$, and low to moderate initial $e$, usually continue along these orbits uninterrupted for a long time.

For a similar reason, we see that the red histogram peaks to the left of the blue one. It may seem surprising, since relaxation is the dominant process and we may expect complete mixing in phase space, how come then there seems to be some memory? Stars which were accreted from small $a$ (circular orbits) seem to have a tendency to be at smaller $a$ also at $t=0$. This is not a surprise, because relaxation is much less efficient in energy (semi-major axis) space than in angular momentum (eccentricity) space. In eccentricity, we can expect fast mixing within one relaxation time. In fact, the classical definition of $t_\mathrm{rx}$ in standard textbooks relates exactly to this point, that the velocity vector changes by $90^\circ$ or so, it does not relate to energy changes, which only occur in real systems rather than the idealized one discussed by \citet{Chandrasekhar1943}.

\section{Mass growth and eccentricities}
\label{sec:RES}

The bulk of the results from the models with constant disc height ($h=h_z$) listed in Table~\ref{ModelParas} are presented here with the aim of showing the effect of increasing the particle and spatial resolution. The effect of resolution is most evident with regards to the growth of the MBH and the eccentricity distribution of the accreted stars. Note that all models have a fixed gas disc mass, so there is no contribution to the growth of the MBH from gas, only from the accretion of stars.

In addition to the MBH mass growth, the eccentricity distribution of accreted stars is a key focus rather than any other orbital element. This is due to the fact that all accreted stars have (by definition) a pericentre distance $\leq r_\mathrm{acc}$; thus, the semi-major axis is constrained by the eccentricity according to $a \leq r_\mathrm{acc}/(1-e)$. In Section~\ref{sec:hrdisc}, the orbital inclination of the accreted stars is also investigated in the context of different gas disc profiles.

\begin{figure}
\begin{centering}$\begin{array}{c}
\multicolumn{1}{l}{\mbox{(a) MBH growth}}\\[-0.1cm]
\includegraphics[width=\columnwidth]{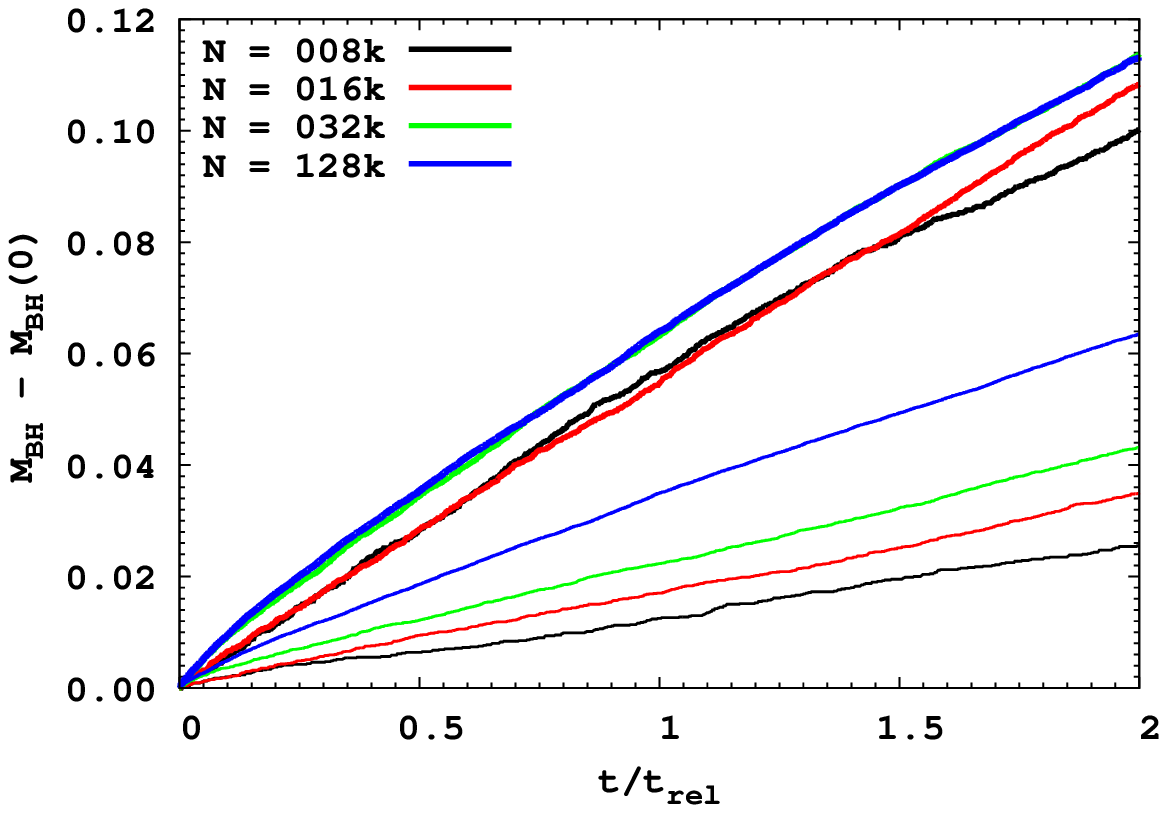}\\
\multicolumn{1}{c}{\mbox{}}\\
\multicolumn{1}{l}{\mbox{(b) Cumulative eccentricity distribution}}\\[-0.1cm]
\includegraphics[width=\columnwidth]{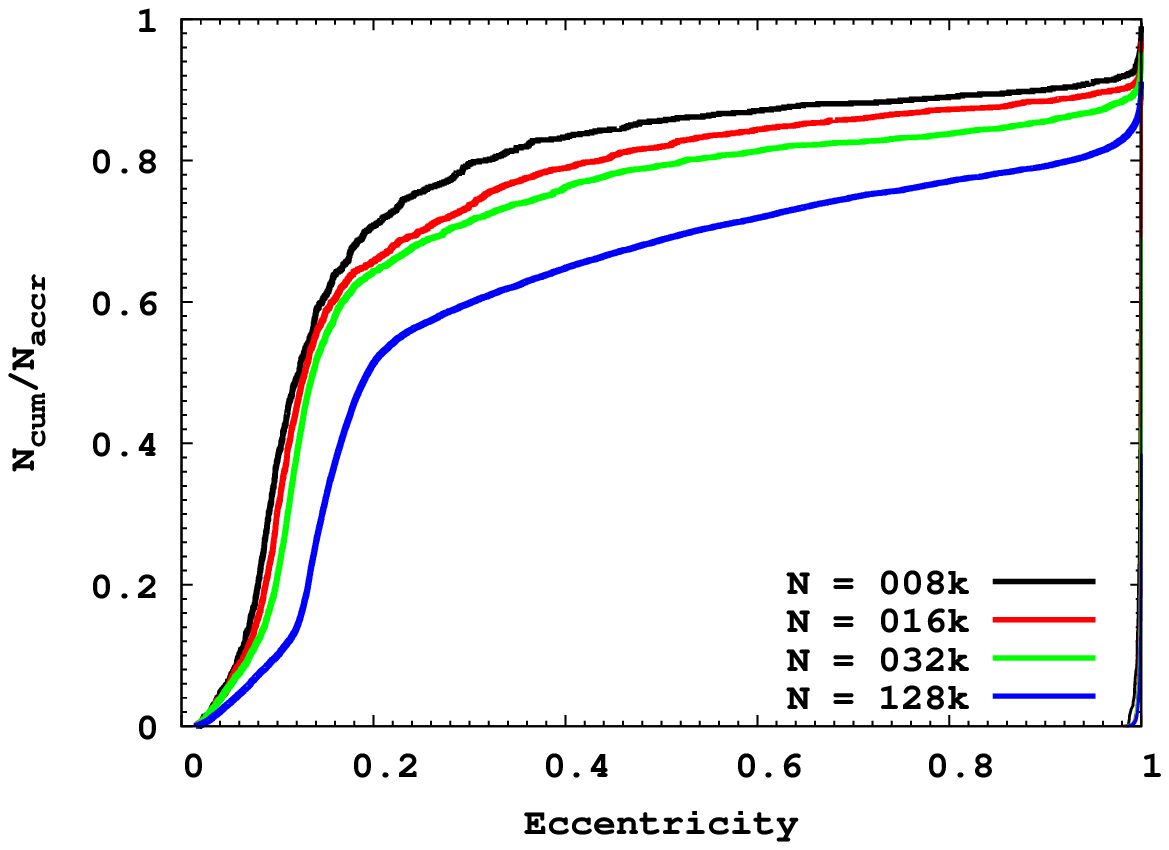}
\end{array}$
\par\end{centering}
\caption{The top panel shows the MBH mass growth with time (in units of the relaxation time) for $r_\mathrm{acc}^* = 3$ with different particle numbers $N$ (denoted by different colours as indicated by the legend). Constant disc thickness models and models without gas are shown as thick and thin lines, respectively; note that the lines for $N=32$k and 128k (with gas disc) overlap. The bottom panel shows the cumulative eccentricity distribution for the same models after two relaxation times (note that the thin lines are difficult to make out, they hug the bottom-right corner of axes, as in this case all accreted particles have very high eccentricities).}
\label{Res:FixedRVaryN}
\end{figure}

The effect of particle resolution on the growth of the MBH is shown in Fig.~\ref{Res:FixedRVaryN} (a). This figure shows the MBH mass versus time in units of the relaxation time, which depends on the total particle number $N$ via equation~(\ref{trelax}). To better illustrate the effect of the particle resolution, the best spatial resolution ($r_\mathrm{acc}^* = 3$) is used throughout. The colours indicate each particle number as stated in the figure caption. Thick lines indicate models with a gas disc and thin lines indicate models with no gas disc (for clarity, this convention is adopted throughout the paper. Note that the mass growth versus time is effectively constant once $N \geq 32$k, i.e. the 32k (blue lines) and 128k (green lines) almost completely overlap in Fig.~\ref{Res:FixedRVaryN} (a).

Fig.~\ref{Res:FixedRVaryN} (b) shows the cumulative eccentricity distribution of accreted stars after two relaxation times. In this figure, $N_\mathrm{accr}$ is the total number of accreted stars after two relaxation times while $N_\mathrm{cum}$ is the number of accreted stars in each with eccentricity $\leq e$. The cumulative eccentricity distribution is then given by $N_\mathrm{cum}/N_\mathrm{accr}$ and is calculated from $e_\mathrm{acc}$, the final eccentricity of the particle when it was removed from the simulation. Note that there is no convergence of the results with increasing particle number. In fact, as the particle number increases, the fraction of stars accreted with low eccentricity falls, i.e. more stars are being accreted with relatively high eccentricity.

\begin{figure}
\begin{centering}$\begin{array}{c}
\multicolumn{1}{l}{\mbox{(a) MBH growth}}\\[-0.1cm]
\includegraphics[width=\columnwidth]{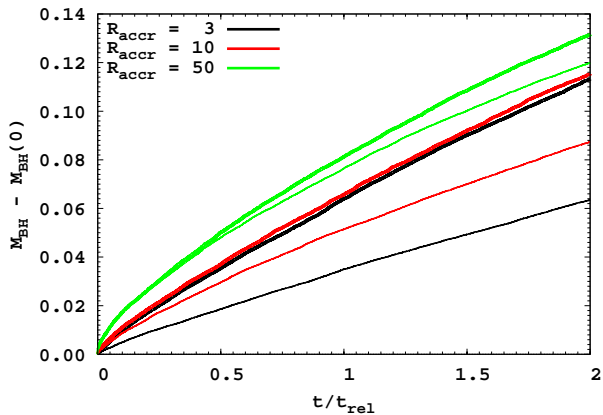}\\
\multicolumn{1}{c}{\mbox{}}\\
\multicolumn{1}{l}{\mbox{(b) Cumulative eccentricity distribution}}\\[-0.1cm]
\includegraphics[width=\columnwidth]{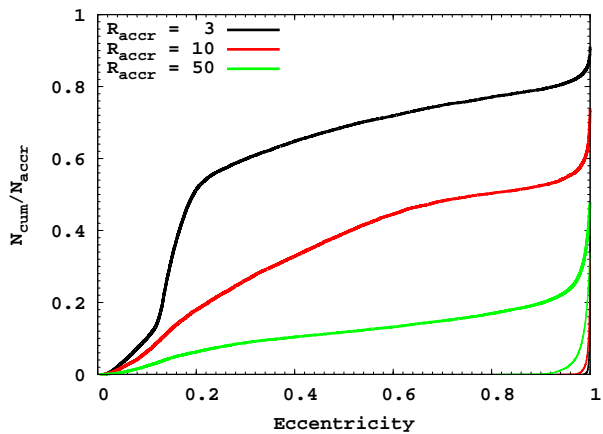}
\end{array}$
\par\end{centering}
\caption{Similar to Fig. \ref{Res:FixedRVaryN} but the particle number is fixed $N=128$k while we vary the spatial resolution, as represented by the numerical accretion radius $r_\mathrm{acc}^* \equiv r_\mathrm{acc}/(10^{-4} R_\mathrm{d})$. Note that the lines for $r_\mathrm{acc}^*=10$ and 3 (with gas disc) overlap in panel (a).}
\label{Res:FixedNVaryR}
\end{figure}

Similar plots for the mass growth and cumulative eccentricity distribution after two relaxation times are shown in Fig.~\ref{Res:FixedNVaryR} for a fixed particle number of $N=128$k and varying the accretion radius to be $r_\mathrm{acc}^* =3,\ 10$ and 50. Fig.~\ref{Res:FixedNVaryR} (a) shows that the mass growth of the MBH does converge as $r_\mathrm{acc}$ decreases for the cases with gas (thick lines), but not for the cases without gas (thin lines). Note that the green line exactly reproduces the result in \PaperI.

\begin{figure}
\includegraphics[width=\columnwidth]{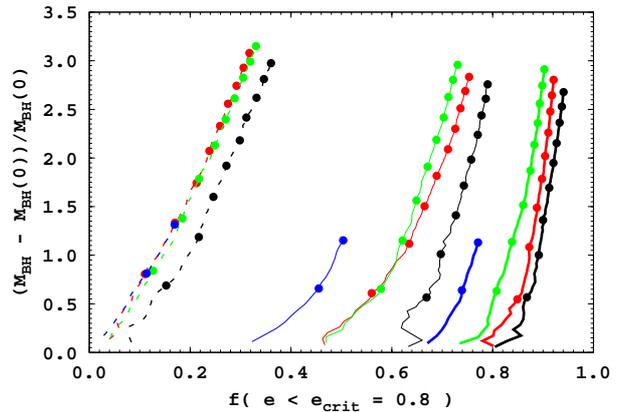}
\caption{The time evolution of all models with constant height gas discs for all 12 combinations of $N$ and $r_\mathrm{acc}^*$. Time increases upwards and filled circles indicate integer relaxation times. Colours indicate the particle resolution with $N=8$k (black), 16k (red), 32k (green) and $N=128$k (blue); line styles indicate the spatial resolution with $r_\mathrm{acc}^*=3$ (thick solid lines), 10 (thin solid lines) and 50 (dashed lines). Note that the $N=128$k runs go only up to 2 relaxation times due to computational cost.}
\label{Res:modelevoln}
\end{figure}

Another way to compare the different models is to show their evolutionary track in a phase space consisting of MBH mass and the fraction $f(e_\mathrm{acc}<e_\mathrm{crit})$ of accreted stars with eccentricity smaller than some value $e_\mathrm{crit}$. We choose $e_\mathrm{crit} = 0.8$ for this critical value as it clearly distinguishes between scenarios where mostly radial orbits are accreted and those where the gas disc dominates. When there is no disc, all accreted stars come from nearly radial orbits, so this fraction $f$ would be zero, otherwise its value carries information about the eccentricity distribution. For the bulk of the simulations with gas, the full time evolution is shown in Fig.~\ref{Res:modelevoln} where all models begin at the bottom and evolve upwards. The filled circles indicate integer relaxation times. Note that the MBH growth slows down (filled circles get closer) for all models as time progresses because of the depletion of stars particles replenishing the disc and MBH loss-cones.

The main results from Fig.~\ref{Res:FixedRVaryN} are that more stars are directly accreted as $N$ increases but increasing the particle resolution beyond $N=32$k does not alter the growth rate of the MBH through star accretion. Increasing the number of stars means that the probability for a star to be scattered into the MBH loss-cone increases and that the loss-cone is kept full for a longer time. This effect must have a steeper-than-linear dependence on $N$, otherwise it would be cancelled out by the linear scaling with $N$ of the 2-body relaxation time. The saturation of the growth rate with $N$ means that the most important physics for capturing stars into the accretion disc and subsequently onto the MBH is adequately captured by only $N=32$k particles. Since the effective loss-cone for capturing stars into the accretion disc is $R_\mathrm{eff} = 0.032$  (see Section~\ref{plungestats}), the diffusion of stars into this region is modelled correctly with relatively small particle numbers as compared to real galactic centres.

From Fig.~\ref{Res:FixedNVaryR} we learn that stars are more easily accreted onto the MBH for larger $r_\mathrm{acc}$ values. This is also clear from Fig.~\ref{Res:modelevoln} and can be understood by considering that models with larger $r_\mathrm{acc}$ have a larger target to hit and so it is easier for the MBH to directly accrete stars. This also explains why the presence of the gas disc is less important for larger values of $r_\mathrm{acc}$, as seen in Fig.~\ref{Res:FixedNVaryR} (b).

This has ramifications for applying these results to real galactic centres. As $r_\mathrm{acc}$ will tend to the stellar tidal disruption radius, from Fig.~\ref{Res:FixedNVaryR} (b) we would expect all orbits to have $e \sim 0$, i.e. $f(e<e_\mathrm{crit}) \sim 1$. However, real galactic centres also have larger particle numbers with $N\sim 10^{9}$ and as noted before, we expect more stars with radial orbits to be accreted with increased particle number. The result of this competition between two effects for real systems is discussed in more detail in Section~\ref{sec:realgc}.

\section{Increasing model realism}
\label{sec:hrdisc}

In this Section we test the effects of two improvements in the code as compared to {\PaperI}: smoothing length for the MBH was removed ($\epsilon_\mathrm{bh}=0$) for the models ending with \texttt{ns} in Table~\ref{ModelParas}, and the disc height was changed from a constant $h_z$ to a two-part function $h(R)$ of equation~(\ref{eq:hR}) to represent the transition to a vertically self gravitating disc.

The scale height profile, which was kept constant in {\PaperI} and Section \ref{sec:RES}, can give incorrect orbits for captured stars when the tidal radius is smaller than the disc scale height. Constant scale height means that there is a cylinder of gas vertically above the MBH, rotating with the planar Kepler velocity; this is unphysical. In this Section we examine the effect of changing the scale height such that it linearly increases with radius, up to the radius $R_\mathrm{sg}$ where the disc becomes vertically self-gravitating given by equation~(\ref{rsg}). For $R > R_\mathrm{sg}$ the scale height is kept constant as $h=h_z=10^{-3}R_\mathrm{d}$ as described in Section~\ref{sec:discmodel}. Note that changing the scale height does not change the total mass of the disc or the surface density at each radius, by consideration of equation~(\ref{gasdensity}).

\begin{figure}
\begin{centering}$\begin{array}{c}
\multicolumn{1}{l}{\mbox{(a) Black hole growth}}\\[-0.1cm]
\includegraphics[width=\columnwidth]{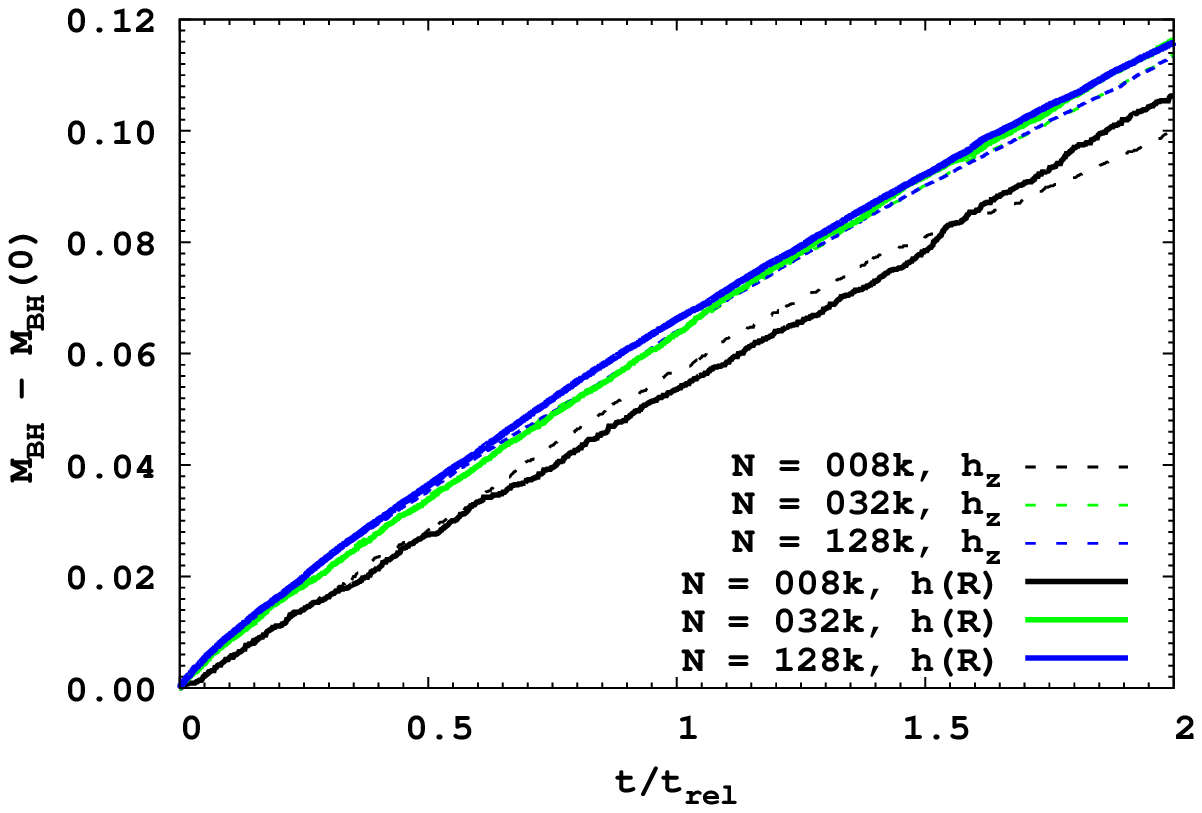}\\
\multicolumn{1}{c}{\mbox{}}\\
\multicolumn{1}{l}{\mbox{(b) Cumulative eccentricity distribution}}\\[-0.1cm]
\includegraphics[width=\columnwidth]{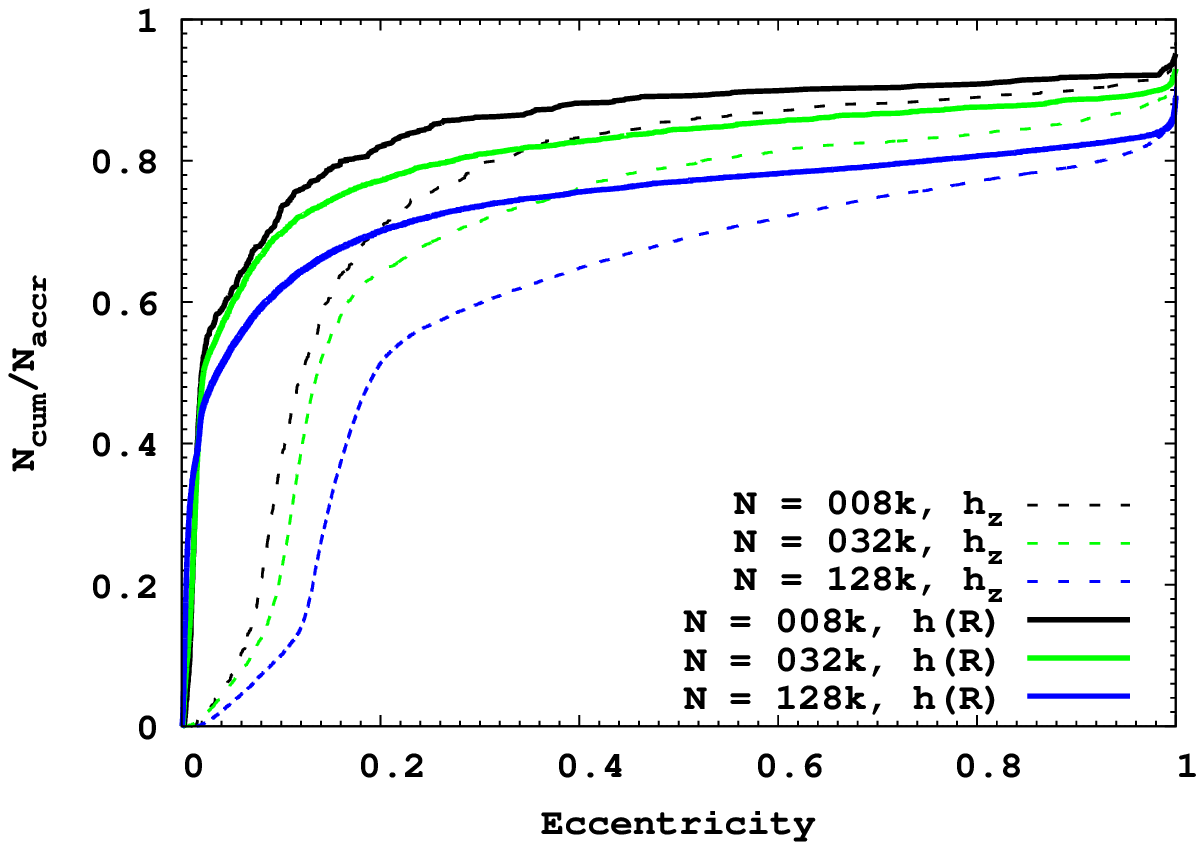}
\end{array}$
\par\end{centering}
\caption{The top panel shows the MBH mass growth for $r_\mathrm{acc}^* = 3$ models for 2 relaxation times for different disc height models. Three particle resolutions were chosen to examine this effect, namely $N=8$k, 32k and 128k.}
\label{Res:DiscHeight}
\end{figure}

Fig.~\ref{Res:DiscHeight} (a) shows that the MBH growth rate is unaffected by the different disc model; the differences we see can be attributed to stochastic variations between numeric models with the same physics. This is expected, since the only difference between the two disc models occurs inside of $R_\mathrm{sg}$, which is inside the effective disc radius $R_\mathrm{eff}$ (determined in Section~\ref{plungestats}). This means that any star close enough to notice the difference between the disc models is already plunging through the accretion disc and onto the MBH.

The only difference in results between the two disc models is in the orbits of the captured stars, as seen in Fig.~\ref{Res:DiscHeight} (b). This Figure shows that the new $h(R)$ disc model leads to a sharper cut off between radial and low eccentricity orbits, i.e. there is a smaller fraction of stars in the eccentricity mid-range. This is true for all particle resolutions. Also note that when $N$ increases, the fraction of stars on radial orbits is also increased, which was seen previously for the constant height disc model in Fig.~\ref{Res:FixedRVaryN} (b).

\begin{figure}
\begin{centering}$\begin{array}{c}
\multicolumn{1}{l}{\mbox{(a) $h=h_z$}}\\
\includegraphics[width=\columnwidth]{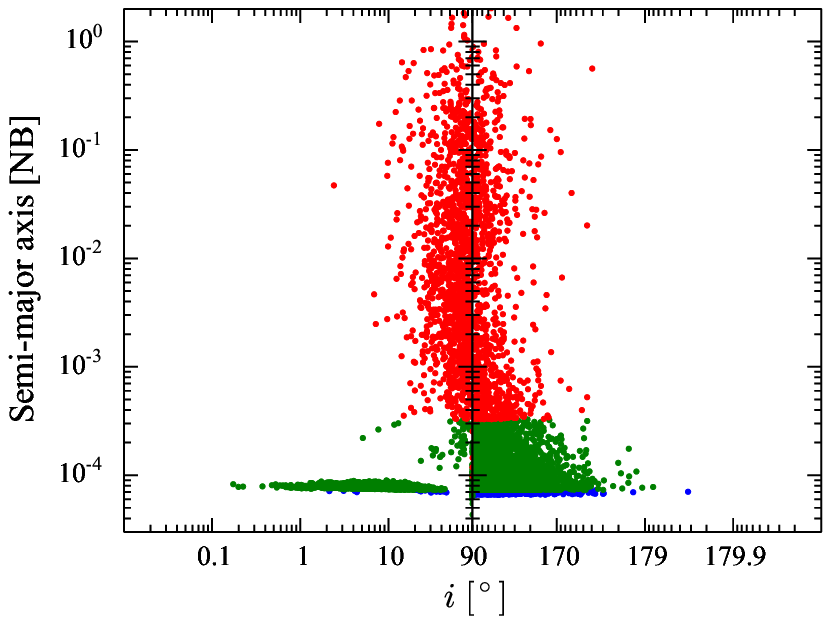}\\
\multicolumn{1}{c}{\mbox{}}\\
\multicolumn{1}{l}{\mbox{(b) $h=h(R)$}}\\
\includegraphics[width=\columnwidth]{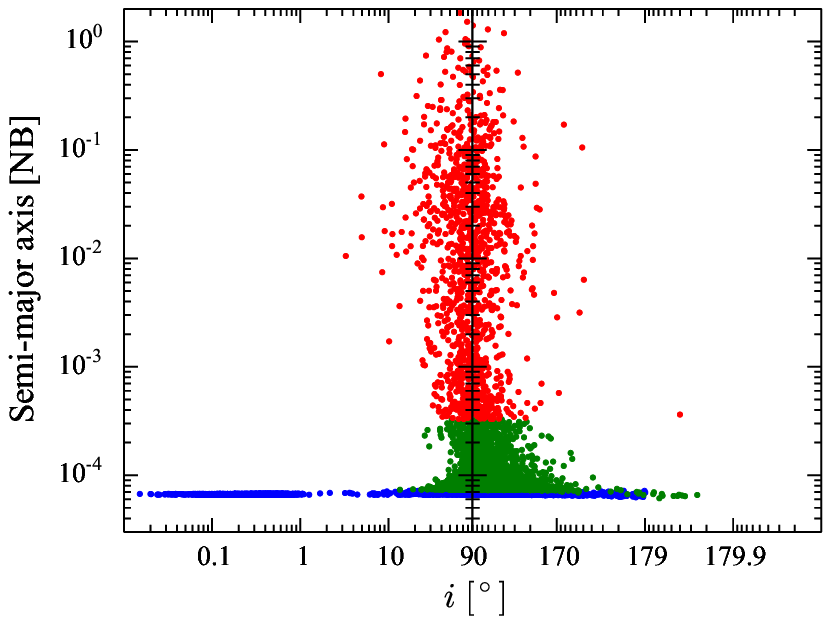}
\end{array}$
\par\end{centering}
\caption{Semi-major axis and orbital inclination for all accreted stars at the moment of accretion for models with constant height disc (panel a; \texttt{128k03}) and varying disc heigh (panel b; \texttt{128k03r}). Particles are divided into three (final) eccentricity groups and are coloured accordingly: $e_\mathrm{acc}>0.8$ (red), $0.1<e_\mathrm{acc}<0.8$ (green) and $e_\mathrm{acc}<0.1$ (blue). Note the non-standard horizontal axis which is used to highlight the differences between particles captured on prograde and retrograde orbits.}
\label{Res:AccrHR}
\end{figure}

To show the effect of the disc model on the orbits of the captured stars in greater detail, a high spatial and particle resolution simulation was performed, namely \texttt{128k03r} in Table~\ref{ModelParas}. This model was previously used in Section~\ref{plungestats} to produce statistics for the radius at which stars begin to plunge into the disc and the time taken for this process. Here we focus on the differences between accreted star orbits for $h_z$ and $h(R)$ models, having already established that mass growth is unaffected. Fig.~\ref{Res:AccrHR} shows the semi-major axis and orbital inclination for all accreted stars at the moment of accretion for models \texttt{128k03} (panel a) and \texttt{128k03r} (panel b). Note that the horizontal axis is composed of two log scales, both ending at $90^\circ$, with one beginning near zero (left) and the other beginning at $180^\circ$ (right). This non-standard plotting method highlights the differences between stars captured on prograde ($i < 90^\circ$) and retrograde ($i > 90^\circ$) orbits.

The stars accreted from radial orbits show very similar characteristics in both disc models in Fig.~\ref{Res:AccrHR} (red dots), in particular the semi-major axis distribution. This is expected, since most radial accretions should originate at the maximum of the stellar density profile, which is the same independent of disc model. There are two key differences between the accreted star populations in the different disc models. First, the population of disc-captured low-eccentricity and prograde-planar stars ($e\sim0$, $i<1^\circ$) is more well-defined for the $h(R)$ model; second, the $h_z$ model has very few stars accreted from inclined ($i>30^\circ$), prograde orbits with semi-major axes close to $r_\mathrm{acc}$. The second point can be understood by removing the unphysical gas above the MBH and inside of $R_\mathrm{sg}$, causing these orbits to rapidly decay in a violent way before they are properly aligned.

Violent or chaotic accretion is seen for retrograde orbits in both disc models where the backwind effect (described in Section~\ref{sec:plunge}) causes the orbit to decay faster than it is aligned with the disc. This behaviour was seen for the plunges shown in Fig.~\ref{FigPlunges} where the star particle either has time to circularise before accretion, panel (b), or is accreted before circularisation, panel (c). For the $h(R)$ disc far more stars are circularised and aligned in the disc (Fig.~\ref{FigPlunges} a) than for the $h_z$ disc. The $h(R)$ model also cause accretion of stars which are aligned in a retrograde sense but have not circularised (green points near $i=179^\circ$), which the constant height model does not.

We note here that we use the word `circularised' for all particles with low $e_\mathrm{acc}$ since all accreted particles, i.e. having pericentre distance $\lesssim r_\mathrm{acc}$, are \textit{initially} on nearly radial orbits in the star cluster. So if they end up with low eccentricity just before accretion, it is because they have been circularised by interactions with the disc. Star--star interactions can refill the loss-cone, but only with particles nearly radial orbits as is evident from the no-gas results.

In both models the number of accreted stars from radial orbits is similar: 22 per cent for $h_z$ and 19 percent for $h(R)$. Far more stars are circularised in the $h(R)$ model compared to the $h_z$ model: 63 percent and 10 percent, respectively. Overall, the more physical $h(R)$ model better separates the radially accreted population and the disc captured population with only 19 percent having $0.1<e_\mathrm{acc}<0.8$ compared to 67 per cent for $h_z$. A more detailed breakdown of the $h(R)$ model, examining the statistics of the plunge and comparing the types of plunges was presented in Section~\ref{plungestats}.

\section{Application to real galactic centres}
\label{sec:realgc}

The number of particles required to accurately model a galactic centre using a direct \textit{N}-body simulation is well beyond current computing capabilities. To apply our results to real galactic centres, we first compare the accretion rate expected from loss-cone theory with the results of the no-gas simulations (see Table \ref{RealEst}), thus ensuring that our models are broadly correct. Then, we extrapolate the results of our models which do have a gas disc to estimate MBH mass growth due to star accretion in real galactic centres.

\subsection{Without gas}
\label{sec:losscone}

In this Section we use loss-cone theory \citep{FR1976} to derive
a formula for the accretion or tidal disruption rate as a function of free parameters,
and in particular we are interested in the scaling with $r_{\mathrm{acc}}$
and $N$. We will then show that our runs without a gas disc scale
roughly in agreement with the formula. Let us assume the MBH to be
a stationary target of radius $r_{\mathrm{acc}}$ and that $N$ particles
form a \citet{BahcallWolf1976} cusp around it, where the
stellar density runs with $r^{-7/4}$, and that the only process
that can cause particles to walk in energy-angular momentum space
is 2-body relaxation. The definition of the loss-cone then is: the
region in angular momentum and energy phase space where stars on 
unperturbed orbits in the given gravitational potential have
an orbital pericentre distance to the MBH less than the accretion radius 
$r_{\mathrm{acc}}$. The loss cone ansatz has been used many times to
estimate black hole growth rates
\citep[cf.][]{LightmanShapiro1977,PauEtAl2004,BaumgardtEtAl2004,ZhongEtAl2014}.
Its main assumption is that the bulk of the tidally accreted stars originates
from a critical radius $r_{\mathrm{crit}}$, where draining of the loss
cone through tidal accretion and refilling due to local 2-body relaxation
balance each other. The MBH accretion rate derived from loss cone
theory is denoted here with $\dot{M}_{\mathrm{bh}}^{\prime(\mathrm{stars})} $
to distinguish it from the value measured in our simulations; the superscript indicates that the change of
mass is due to accretion of stars (rather than gas). It is dominated by the slowest
process and can be estimated through
\begin{equation}
\dot{M}_{\mathrm{bh}}^{\prime(\mathrm{stars})} = \frac{\rho(r_{\mathrm{crit}})r_{\mathrm{crit}}^{3}}{t_{\mathrm{rx}}(r_{\mathrm{crit}})}\label{eqn:mdot}
\end{equation}
Note that close 2-body encounters in the cusp, which kick stars away, may differ
only by a logarithmic factor from the flux of stars towards the black hole
\citep{LinTremaine1980}. However, since usually good agreement between loss cone
models and $N$-body simulations is found, we conclude that this process may be 
neglected at least with regard to the MBH growth. Here we follow a slightly
different ansatz of \citet{PauEtAl2004} where no specific model is assumed
for the density distribution of the central star cluster:
\begin{equation}
\frac{r_{\mathrm{crit}}}{r_{\mathrm{acc}}} = \frac{4t_{\mathrm{rx}}(r_{\mathrm{crit}})}{t_{\mathrm{dyn}}(r_{\mathrm{crit}})}.\label{eqn:rc}
\end{equation}
Here, we have used the conditions at the critical radius as given in equations (11) to (14) of
\citet{PauEtAl2004}.

\begin{figure}
\begin{centering}
\includegraphics[width=1\columnwidth]{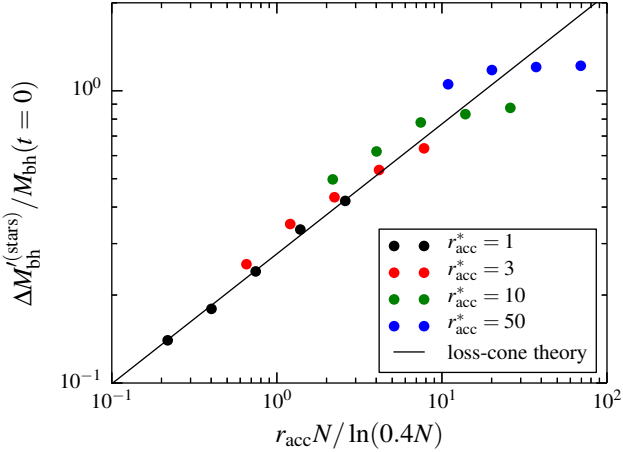} 
\par\end{centering}

\protect\caption{Mass growth of the MBH relative to its initial mass after two relaxation times for all
models without a gas disc (i.e. the evolution is purely stellar dynamics);
the horizontal axis is a combination of the numerical accretion radius
$r_{\mathrm{acc}}$ and particle number $N$. The solid line shows
the best-fitting power-law with index of $4/9$ to the models with
$r_{\mathrm{acc}}^{*}=1$ only (filled black circles), as expected
from loss-cone theory (equation \ref{eqn:delMth}). Other coloured
circles represent different values of $r_{\mathrm{acc}}^{*}$ as indicated
by the legend.}

\label{fig:DeltaMng} 
\end{figure}

We now rescale the dynamical and relaxation time scales to the influence radius of the
MBH $r_{\mathrm{inf}} = G M_{BH}/\sigma^2$, assuming $r_{\mathrm{crit}} < r_{\mathrm{inf}}$ and using
$\sigma \propto r^{-1/2}$ for the velocity dispersion in the stellar system dominated
by the MBH gravity (i.e. $r < r_{\mathrm{inf}}$). The stellar density inside the influence radius is taken as the Bahcall Wolf cusp scaled to the density at the influence radius, given by
\begin{equation}
\rho(r) = \rho(r_{\mathrm{inf}}) \left( \frac{r_{\mathrm{inf}}}{r} \right)^{7/4}.
\end{equation}
Substituting these into the following expressions for the relaxation time and the dynamical time at $r=r_{\mathrm{crit}}$ 
\begin{eqnarray}
t_{\mathrm{rx}}(r_{\mathrm{inf}}) & \propto &  \sigma(r_{\mathrm{inf}})^{3}/[\rho(r_{\mathrm{inf}}) m_\star \ln\Lambda ]\label{eq:trx-rc}\\
t_{\mathrm{dyn}}(r_{\mathrm{inf}}) & \propto & r_{\mathrm{inf}}/\sigma(r_{\mathrm{inf}}).\label{eq:tdyn-rc}
\end{eqnarray}
with $m_\star = 1/N$ and noting that $r_{\mathrm{inf}} \propto M_{BH}$ we get 
\begin{equation}
r_{\mathrm{crit}}^{9/4} = \zeta r_{\mathrm{acc}} \frac{N}{\ln\Lambda} M_{\mathrm{bh}}^{1/4}
\end{equation}
Here we have collected in the constant $\zeta$ all other quantities, which do {\em not} depend
on $r_{\mathrm{acc}}$, $N$, or $M_{\mathrm{bh}}$, which are the quantities relevant for our
discussion. $\zeta$ will vary depending on the stellar system quantities at $r_{\mathrm{inf}}$ and
some physical constants; it is not dimensionless. With this we get the relation
\begin{equation}
r_{\mathrm{crit}} \propto \left[ r_{\mathrm{acc}} \frac{N}{\ln\Lambda} \right]^{4/9} M_{\mathrm{bh}}^{1/9}
\end{equation}
The experimental quantity we measure from the simulations in our numerical experiments (and would like to compare to theory) is the mass growth per fixed time interval (here two relaxation times at the half-mass radius), which we denote as $\Delta M_\mathrm{bh} = 2 t_\mathrm{rx}(r_\mathrm{h}) \dot{M}_{\mathrm{bh}}$; so we get a final scaling of
\begin{equation}
\label{eqn:delMth}
\frac{\Delta M_\mathrm{bh}}{M_\mathrm{bh}} = k \left[ r_{\mathrm{acc}} \frac{N}{\ln(0.4 N)} \right]^{4/9} M_{\mathrm{bh}}^{10/9}.
\end{equation}
In the equation above we have identified the Coulomb logarithm at the critical radius $\ln\Lambda$ with the Coulomb logarithm $\ln (0.4N)$ occurring in the half-mass relaxation time, as in {\PaperI}. We compare equation (\ref{eqn:delMth}) directly with our numerical results from the
models without gas as shown in Fig.~\ref{fig:DeltaMng}. Since we
expect models with smaller $r_{\mathrm{acc}}$ to give a better approximation
to the loss-cone theory, we ran an extra set of models (without a
gas disc) with $r_{\mathrm{acc}}^{*}=1$ for each $N$ (filled black
circles). The models with larger $r_{\mathrm{acc}}^{*}$ indeed show
a large deviation from the theoretical expectations. Notice that small
deviations from the theoretical curve may depend either on $M_{\mathrm{bh}}$
or on the quantities from outside the MBH influence radius absorbed in the
constant $k$ above.

A small accretion radius for the MBH and no gas disc should have the
same power-law relationship as equation~(\ref{eqn:delMth}). We put
it to the test by fitting a power-law to the five data-points with
$r_{\mathrm{acc}}^{*}=1$. The best fitting power-law index was $\approx0.456$,
which is within 3 per cent from the theoretical value of $4/9$. To
extrapolate our results to real galactic centres in this case, we
force the index to be $4/9$ and find the best-fitting constant, we
get $k\approx0.273$. This fit is shown as the solid black line in Fig.~\ref{fig:DeltaMng}. 

\subsection{With gas}

\label{sec:realgcs}

The previous section established that simulation results without gas are consistent with theoretical expectations, giving us confidence in the numerical code. Here, the results from models with a gas disc are extrapolated into the regime of real galactic centres. The highest resolution and most realistic model with a gas disc, \texttt{128k03r}, had $N=128$k, $r_{\mathrm{acc}}^{*}=3$ and $h(R)$ disc height profile. The MBH mass growth as a function of time for this model is shown as the blue curve in Fig.~\ref{Res:DiscHeight} (a) for two relaxation times. We found that we could fit this curve using a simple power-law given by
\begin{equation}
g(t)=a\left(\frac{t}{t_{\mathrm{rx}}}\right)^{b}\label{mgrowthfit}
\end{equation}
where $a\approx0.0643$, $b\approx0.843$ and $t_{\mathrm{rx}}$ is the relaxation time for $N=128$k. The residuals were less than 0.01 mass units, i.e. better than a 10 per cent error relative to the initial MBH mass.

\begin{table*}
\caption{Estimation of mass growth via accretion of stars and gas for a sample of galactic nuclei.}
\label{RealEst}
\begin{tabular}{lllllll}
\hline
Object & $M_\mathrm{bh}$ & $r_\mathrm{inf}$ & $\dot{M}_\mathrm{bh}^{\prime(\mathrm{stars})}$ & $\dot{M}_\mathrm{bh}^{(\mathrm{stars})}$ & $\dot{M}_\mathrm{bh}^{(\mathrm{gas})}$ & $\dot{M}_\mathrm{Edd}$\\
 & ($\msun$) & (pc) & ($\msun\,\mathrm{yr}^{-1}$) & ($\msun\,\mathrm{yr}^{-1}$) & ($\msun\,\mathrm{yr}^{-1}$) & ($\msun\,\mathrm{yr}^{-1}$)\\
\hline
M 87       & $6.6 \times 10^{9}$ &  291 & $4 \times 10^{-5}$ & $9 \times 10^{-5}$ & $1 \times 10^{-2}$ & $1 \times 10^{2}$\\
NGC 3115   & $9.6 \times 10^{8}$ &   78 & $3 \times 10^{-5}$ & $1 \times 10^{-4}$ & $2 \times 10^{-2}$ & $2 \times 10^{1}$\\
NGC 4291   & $3.2 \times 10^{8}$ &   24 & $7 \times 10^{-5}$ & $3 \times 10^{-4}$ & $5 \times 10^{-2}$ & $7$\\
M 31       & $1.5 \times 10^{8}$ &   25 & $2 \times 10^{-5}$ & $2 \times 10^{-4}$ & $2 \times 10^{-2}$ & $3$\\
NGC 4486A  & $1.3 \times 10^{7}$ &  4.5 & $4 \times 10^{-5}$ & $3 \times 10^{-4}$ & $3 \times 10^{-2}$ & $3 \times 10^{-1}$\\
MW         & $4.0 \times 10^{6}$ &  1.4 & $1 \times 10^{-4}$ & $7 \times 10^{-4}$ & $7 \times 10^{-2}$ & $9 \times 10^{-2}$\\
M 32       & $3.0 \times 10^{6}$ &  2.3 & $3 \times 10^{-5}$ & $3 \times 10^{-4}$ & $2 \times 10^{-2}$ & $7 \times 10^{-2}$\\
\hline
\end{tabular}
\par\medskip
\begin{flushleft}\textbf{Notes.} We extrapolate results to this sample of galactic nuclei (adopted from {\PaperI}). Columns 1--3 are the object's name, MBH mass and radius of influence (calculated from the stellar velocity dispersion), respectively. Column 4 gives the expected accretion rate of stars given pure stellar dynamical evolution (no gas disc); Column 5 gives the expected accretion rate of stars assuming a gas disc of the form assumed in this work is present (averaged over 100\,Myr); Column 6 gives the expected accretion rate of gas from said disc model; Column 7 is the accretion rate corresponding to the Eddington luminosity for each MBH, assuming 10 per cent accretion efficiency.\end{flushleft}
\end{table*}

In order to extrapolate from the simulation results, a number of assumptions are made: (1) mass ratios between disc, MBH and nuclear stellar cluster are roughly the constant adopted in Section \ref{sec:method}, (2) the time evolution of the mass ratios does not significantly affect the MBH mass growth, (3) we only need to extraplolate to a larger $N$, but the value $r_{\mathrm{acc}}^{*}=3$ used in our models is already small enough as justified in Section \ref{sec:RES}, and (4) the disc can be described by a thin $\alpha$-disc with gas on Keplerian orbits. These assumptions are clearly flawed to some extent, but they do allow us to perform an order of magnitude calculation. A more realistic model of the galactic centre is our long term goal.

Table~\ref{RealEst} shows a sample of galactic nuclei, also used for scaling in {\PaperI} along with their MBH masses and influence radii; these are the only two quantities used for scaling. The stellar mass of the cluster $M_\mathrm{cl}$ is assumed to be $\times 10$ that of the MBH, and the number of stars is assumed $N=M_\mathrm{cl}/\msun$. The disc radius $R_\mathrm{d}$ is assumed $=r_\mathrm{inf}$ and the cluster half-mass radius $r_\mathrm{h}$ is assumed $=3R_\mathrm{d}$. We need to use a physical accretion radius when scaling equation~(\ref{eqn:delMth}) to real galaxies. For the four lowest MBH masses in the Table, we approximate it with the rigid-body Roche limit for a Sun-like star
\begin{equation}
r_\mathrm{acc} = \rsun \left( 2 M_\mathrm{bh}/\msun \right)^{1/3},
\end{equation}
while for the three highest masses this turns out to be smaller than the Schwarzschild radius, so accretion will take place without tidal disruption, and for the scaling of equation~(\ref{eqn:delMth}) we use
\begin{equation}
r_\mathrm{acc} = 2 G M_\mathrm{bh}/c^2.
\end{equation}
Note that it is only used to calculate $\dot{M}_\mathrm{bh}^{\prime(\mathrm{stars})}$, the other quantities in the Table are agnostic to the real accretion radius. The extrapolation of $\dot{M}_\mathrm{bh}^{(\mathrm{stars})}$ is straightforward using equation~(\ref{mgrowthfit}), one just needs to find the relaxation time at the half-mass radius for the given nuclei using equation~(\ref{trelax}) and convert to physical mass units by multiplying $g(t)$ by $M_\mathrm{cl}$. For details on the origin of these values see {\PaperI} and references therein. We use 100\,Myr as the lifetime of the disc \citep[see][]{MillerEtAl2003} and divide the mass growth by this time-scale to get the average accretion rate.

Another contribution to the growth of the MBH is gas from the accretion disc itself. The \citet{SS1973} model assumes a constant gas accretion rate to construct the density structure of the thin disc. Since we assume an ad-hoc density structure, we can work back to find the corresponding gas accretion rate. Using equation~(2.19) of \citet{SS1973} and our equation~(\ref{gasdensity}) we find by algebraic manipulation that
\begin{align}
\dot{M}_\mathrm{bh}^{(\mathrm{gas})} &= 1.8\times10^{-6} \, {\mathrm{M}_{\odot}\,\mathrm{yr}^{-1}} \alpha^{8/7} \left(\frac{M_{\mathrm{d}}}{\mathrm{M}_{\odot}}\right)^{10/7}\nonumber\\
 &\phantom{=} \left(\frac{M_\mathrm{bh}}{\mathrm{M}_{\odot}}\right)^{-5/14}\left(\frac{R_{\mathrm{d}}}{\mathrm{pc}}\right)^{-25/14},
\end{align}
where $\alpha$ is the viscosity parameter which is set to $0.2$ for the purpose of calculating the values of $\dot{M}_\mathrm{bh}^{(\mathrm{gas})}$ in Table \ref{RealEst}.

Overall, the contribution of stars to the growth of the MBH is typically $\sim 100$ times weaker than the growth due to accretion of gas from the disc itself. However, this very simple calculation depends strongly on the thin-disc model while accretion in reality is much more complex, so these results should be treated as approximate at best.

\section{Summary and discussion}
\label{sec:CON}

Using a modified version of the $\upvarphi$\textsc{grape} code, we performed a suite of 39 high-accuracy direct $N$-body simulations of a system containing a star cluster, an MBH and a static gas disc. The mass ratios were $M_\mathrm{bh}/M_\mathrm{cl} = M_\mathrm{d}/M_\mathrm{bh} = 0.1$, the rotation profile of the disc was Keplerian and its density profile derived from the Shakura--Sunyaev thin disc model. We tested two disc models (constant height and variable height) and a large range of particle numbers (from 8k to 128k) and accretion radii (from $\sim 10^{-5}$ to $\sim 10^{-3}$ H\'enon length units); we also looked at the effect of the softening length of the MBH--star interaction. All simulations finished at least two relaxation times.

The disc used in our model represents the inner Keplerian part of the gaseous material surrounding the central AGN engine. It is very difficult to resolve it observationally and there is no standard value known for the mass of such disks in general. From modelling and theory, one can argue about the stability of such disks against self-gravity (Toomre $Q$ analysis) or non-axisymmetric perturbations. According to \citet{ArtymowiczEtAl1993}, $\mu_\mathrm{d} = 0.01$ (where $M_\mathrm{d} = \mu_\mathrm{d} M_\mathrm{bh}$) results in a clearly stable disc, while $\mu_\mathrm{d} = 0.2$ is prone to instability \citep{RoedigEtAl2011}. In {\PaperI} we chose $\mu_\mathrm{d} = 0.1$, and argued that such a disc is already marginally unstable (for $R > 0.26 R_\mathrm{d}$). Such marginally stable disc could still survive the life cycle of quasars (of order $10^8$ years), and gives the maximal dissipative effect on stars \citep{Rauch1995}. The gas reservoir after galaxy mergers outside $R_\mathrm{d}$ (which is also about equal to the gravitational influence radius of the MBH) is, according to many simulations \citep[see e.g.][]{CallegariEtAl2009, ChaponEtAl2013}, much larger, even larger than the black hole mass. Therefore it seems plausible to assume that our AGN disc assumes the maximum stable mass. We choose the value $\mu_\mathrm{d} = 0.1$ for simplicity, because the exact stability boundary can only be determined by much more detailed disc modelling \citep[see e.g.][]{SyerEtAl1991} and depends on parameters which are unknown or poorly defined in our work (e.g. disc viscosity, equation of state).

For the models with a gas disc, the MBH growth due to star accretion converges for sufficiently large particle number ($N \sim 32$k) and small accretion radius ($r_\mathrm{acc}^* \sim 3$). This convergence in MBH growth is not seen for models without gas, precisely as expected from loss-cone theory as explained in Section~\ref{sec:losscone}. There is no convergence in the eccentricity distribution of accreted stars for models with a gas disc due to two competing effects. On the one hand, as the particle number $N$ increases, the fraction of accreted stars on low eccentricity orbits falls (characterised by the fraction $f$ of particles accreted with eccentricity smaller than 0.8). On the other hand, as the spatial resolution increases (i.e. $r_\mathrm{acc}^*$ decreases towards the stellar tidal disruption radius), the fraction $f$ increases as more stars are circularised by the disc before being accreted. It is thus difficult to extrapolate the eccentricity distribution to real galaxies, but from inspection of Figs. \ref{Res:FixedRVaryN} (b) and \ref{Res:DiscHeight} (b) it seems that a large fraction of stars will be disrupted when their orbit has been fully circularised (i.e. $e_\mathrm{acc}\sim 0$). This has ramifications for the search of electromagnetic counterparts of gravitational wave emitters.

The improved variable height disc model resulted in a significantly higher fraction of stars that had been circularised before being accreted, compared to the model with constant disc height. There was no difference in the growth of the MBH, however. Data from our most realistic simulation using the improved disc model ($N=128$k, $r_\mathrm{acc}^* = 3$) were used to identify the three main paths whereby stars accreted onto the MBH: \textit{disc capture} ($e_\mathrm{acc}\sim 0$), \textit{gas assisted accretion} ($e_\mathrm{acc}<1$) and \textit{direct accretion} ($e_\mathrm{acc}\sim 1$); these were discussed in detail in Section~\ref{plungestats} along with the statistics of the different types. Using these data we also calculated the effective radius in the disc where stars began to plunge, which was $R_\mathrm{eff} = 0.032$ ($\sim 0.15$ per cent of the disc radius $R_\mathrm{d}$) and the time-scale for this to occur was between 1 and 10 per cent of the 2-body relaxation time at the half-mass radius.

When extrapolating our results to real galactic nuclei, we found that typically the stellar accretion rate is increased by a factor of $\sim 10$ in the presence of the disc, and is typically a few times $10^{-4}\,\msun\,\mathrm{yr}^{-1}$. This assumes that 2-body relaxation is the main process that replenishes the loss-cone, and that our most realistic model with the disc is convergent in terms of both $N$ and $r_\mathrm{acc}$. The former assumption is not justified when the system is strongly non-spherical or massive perturbers are present, both cause large scale torques that can redistribute angular momentum in the stellar system. The latter assumption should be made carefully, since while it is clear that our results for the accretion rate converge between $N=32$k and 128k and between $r_\mathrm{acc}^*=10$ and 3, a realistic system is still a few orders of magnitude away in both variables.

We estimated the accretion rate of gas directly from the disc using the assumptions of the Shakura--Sunyaev model, and found that it is typically $\sim 100$ times larger than the accretion rate of stars, which means that star accretion is not a major contributor to MBH mass growth \citep[cf.][]{MK2005,KennedyEtAl2011}. But while the star accretion rate seems to be robust given the density profile and mass of the disc, the estimated gas rate is potentially much more model-dependent and could be very different, e.g. due to winds. Moreover, our assumption that all the material from the tidally disrupted star is added to the MBH mass is clearly off by some factor.

Since many stars spend some time captured within the disc as they migrate inwards under the effect of friction, if the gas were to suddenly disappear, it would leave behind a disc of stars. The Galactic centre is known to have one \citep{LevinBeloborodov2003} or two \citep{PaumardEtAl2006} such stellar discs, but whether or not this observation is consistent with the subsystem of trapped stars we find in our models will be examined in future work. We note that other mechanism can form a disc-shaped stellar subsystem such as in-situ star formation in the gas disc and vector resonant relaxation \citep{KocsisTremaine2015}. Different mechanisms will have different signatures in terms of mass and age segregation, which cannot be distinguished in our models where all stars are identical and no stellar evolution is included. Multiple stellar populations as well as other realistic physics such as massive perturbers and dynamic disc model will be included in future work.

\citet{ArcaviEtAl2014} find that in their archival search of the Palomar Transient Factory, all three candidate TDEs are in E+A galaxies, that have spectra typical of the host galaxies of type 2 Seyferts \citep{DresslerGunn1983}. Since these galaxies are rare in the local universe, comprising just 0.1 per cent \citep{Goto2007,SnyderEtAl2011}, there seems to be an enhancement of the TDE rate in these galaxies. \citet{ArcaviEtAl2014} speculate that if E+A galaxies are indeed the result of a merger \citep{ZabludoffEtAl1996}, then the TDE rate could be enhanced by the binary MBH mechanism \citep{ChenEtAl2011}. However, because of the similarity of the spectra to AGN hosts, we postulate that a gaseous disc may be present and is responsible to TDE rate enhancement as described in this work. \citet{LiEtAl2015} show in their simulation how the TDE rate increases due to mergers, however, the duration of the event is short; an estimate how this affects TDE rates in a cosmological background is not yet done.

\section*{Acknowledgements}

We thank Maxim Makukov, Chingis Omarov, Emmanuil Y. Vilkoviskij and Ari Laor for valuable comments, discussion, support and supervision. We also thank the anonymous referee for useful comments and suggestions. This work has benefited very much from funding of exchange and collaboration between Germany and Kazakhstan by Volkswagen Foundation under the project ``STARDISK -- Simulating Dense Star-Gas Systems in Galactic Nuclei using special hardware'' (\textit{I/81 396}). GRAPE and GPU hardware at Fesenkov Astrophysical Institute in Kazakhstan used for this work have been supported by the STARDISK project, too. We acknowledge support from the Strategic Priority Research Program ``The Emergence of Cosmological Structure'' of the Chinese Academy of Sciences (No. \textit{XDB09000000}) (Pilot B programme).

We acknowledge support by Chinese Academy of Sciences through the Silk Road Project at NAOC, through the Chinese Academy of Sciences Visiting Professorship for Senior International Scientists, Grant Number \textit{2009S1-5} (RS), and through the ``Qianren'' special foreign experts program of China. RS has also been partially supported by NSFC (National Natural Science Foundation of China), grant Nr. \textit{11073025}. GK acknowledges a Chinese Academy of Sciences Young International Postdoctoral Fellowship Grant No. \textit{2011Y2JB09}. YM acknowledges support from the China Postdoctoral Science Foundation through grant No. \textit{2015T80011}, from the excellence initiative at the Univ. of Heidelberg through mobility measures for international research collaborations project nr. \textit{7.1.47}, and the European Research Council under the European Union's Horizon 2020 Programme, ERC-2014-STG grant GalNUC 638435. PB acknowledges the special support by the NASU under the Main Astronomical Observatory GRID/GPU computing cluster project.

The special GPU accelerated supercomputer {\tt laohu} at the Center of Information and Computing at National Astronomical Observatories, Chinese Academy of Sciences, funded by Ministry of Finance of People's Republic of China under the grant \textit{ZDYZ2008-2}, has been used for some of the largest simulations. We also used smaller GPU clusters {\tt titan}, {\tt hydra} and {\tt kepler}, funded under the grants \textit{I/80041-043} and \textit{I/84678/84680} of the Volkswagen Foundation and grants \textit{823.219-439/30} and \textit{/36} of the Ministry of Science, Research and the Arts of Baden-W\"urttemberg, Germany.

\bibliography{SDRefs}

\end{document}